%% file: main.tex
\documentclass[letterpaper]{article} 
\usepackage{aaai25}  
\usepackage{times}  
\usepackage{helvet}  
\usepackage{courier}  
\usepackage[hyphens]{url}  
\usepackage{graphicx} 
\usepackage{natbib}  
\usepackage{caption} 

\usepackage{longtable}
\usepackage{algorithm}
\usepackage{algorithmic}
\usepackage{newfloat}
\usepackage{listings}
\usepackage{tabularx}
\usepackage{makecell}
\newcommand{\del}[1]{}
\DeclareCaptionStyle{ruled}{labelfont=normalfont,labelsep=colon,strut=off}
\title{Inconsistencies in Classification of Online News Articles:\\A Call for Common Standards in Brand Safety Services}

\author {
    Michael Smith\textsuperscript{\rm 1},
    Riley Grossman\textsuperscript{\rm 1},
    Antonio Torres-Agüero\textsuperscript{\rm 2},
    Pritam Sen\textsuperscript{\rm 1},
    Cristian Borcea\textsuperscript{\rm 1},
    Yi~Chen\textsuperscript{\rm 1}
}
\affiliations {
    \textsuperscript{\rm 1}New Jersey Institute of Technology, Newark, New Jersey, USA\\
    \textsuperscript{\rm 2}DeepSee.io, Los Angeles, California, USA\\
    \{mes6, rag24, ps37, borcea, yi.chen\}@njit.edu, antonio@deepsee.io
}
\providecommand{\yc}[1]{}      
\providecommand{\del}[1]{}      
\providecommand{\rev}[1]{#1}    
\nocopyright
\begin{document}

\maketitle

\begin{abstract}
This study examines inconsistencies in the brand safety classifications of online news articles by analyzing ratings from three leading brand safety providers, DoubleVerify, Integral Ad Science, and Oracle. We focus on news content because of its central role in public discourse and the significant financial consequences of unsafe classifications in a sector that is already underserved by digital ad spending. By collecting data from 4,352 news articles on 51 domains, our analysis shows that brand safety services often produce conflicting classifications, with significant discrepancies between providers. These inconsistencies can have harmful consequences for both advertisers and publishers, leading to misplaced advertising spending and revenue losses. This research provides critical insights into the shortcomings of the current brand safety landscape. We argue for a standardized and transparent brand safety system to mitigate the harmful effects of the current system on the digital advertising ecosystem.
\end{abstract}


\input{introduction}

\input{related}
\input{method}

\input{results}
\input{discussion}

\input{conclusion}
\input{acknowledgements}

\bibliography{aaai25}
\appendix
\input{appendix}

\end{document}

%% file: introduction.tex
\section{Introduction}


\label{sec:intro}

Brand safety services are essential for advertisers to ensure that their ads do not appear alongside harmful or inappropriate content. Their value is best demonstrated in a recent paper that surveyed over 200 decision makers from companies spending more than \$10 million annually on advertising, finding that more than 80\% were concerned with brand safety. These same participants indicated that they would pay a nearly 50\% premium to guarantee that their advertisements would only appear alongside safe content~\cite{johnson2023}, thus highlighting the need for brand safety classification tools.

Our study focuses on DoubleVerify (DV), Integral Ad Science (IAS) and Oracle, because they are three of the most widely used state-of-the-art providers of brand safety technology in the industry~\cite{thetechylife,businessinsider2024,usgov}. 
%
 \del{Oracle emerged as one of the most prominent providers, while IAS and DV were the original pioneers in this space~\cite{businessinsider2024}, securing partnerships with Facebook, Instagram, TikTok, and Snap~\cite{adexchanger2024iasdv}.} Each brand safety provider offers a different, proprietary classification system to assess the safety of online content. But all claim to leverage AI, natural language processing (NLP) and machine learning (ML) technologies to evaluate whether a webpage's content is safe for advertising based on predefined safety categories (e.g., violence, adult, alcohol)~\cite{vargo2024,griffin2023}.





Correct classification of brand safety is critical for both advertisers and publishers, and yet, remains a huge challenge. 
On one hand, an article with safe content may be misclassified as unsafe. For example, an article discussing recent advances in breast cancer research can be classified as unsafe due to the presence of the keyword ``breast''. In such cases, advertisers lose opportunities and publishers face reduced revenues. On the other hand, an article with potentially unsafe content may be misclassified as safe, which results in advertisers wasting their budget and also risking their brand's reputation. \del{ We find that the classification challenge is particularly pronounced and may have a significant impact on breaking news content, which often addresses sensitive topics such as violence or crime, leading to a higher likelihood of being assessed as unsafe compared to other types of content.  }  One well-known example is YouTube's ``Adpocalypse". Despite YouTube's keyword-based brand safety controls, major brands like AT\&T, Pepsi, and Verizon pulled their ads from the platform after discovering their ads were placed alongside extremist or inappropriate content~\cite{guardian2017}.  This highlights how misclassification can lead to financial and reputational damage for advertisers and publishers.

While professional editorial review reduces the likelihood of toxic language compared to user-generated platforms, it does not eliminate risk. News outlets regularly cover violence, pandemics, political extremism, and other sensitive topics that can trigger brand safety systems. In these cases, the underlying content is not ‘unsafe’ in the sense of misinformation or hate speech, but it is nevertheless subject to unsafe ratings due to its subject matter. Thus, news content presents a unique challenge: the coexistence of editorial standards with recurring coverage of high-risk topics. Misclassifying safe digital news content as unsafe will harm the struggling news publishers. Despite generating \$12 billion in digital advertising revenue in 2019, an industry report estimated that news publishers across the United States failed to generate an additional \$2.8 billion because of erroneous, or overly cautious, keyword-blocking practices that prevented safe content from being monetized~\cite{cheq2020}.

\del{Misclassifying unsafe content as safe can also indirectly harm news publishers as advertisers may reduce their spending to protect their brand's reputation and avoid association with unsafe content.}

\yc{\del{youtube seems irrelevant as it's not about news. maybe revert back to the break news text.}}



\del{Print advertising continues to decline as a revenue source for news publishers, as digital advertising revenue grows~\cite{wan-ifra}. Publishers like The New York Times generate more revenue from digital advertising than from print~\cite{cheq2020}. However, recent developments in AI and (e.g., AI overviews in search engine results) have reduced user traffic to news publishers' websites. A 2025 industry report estimates that traffic to news websites from Google searches fell 7\% in comparison to 2024~\cite{dcn_ai_pubs}. Thus, digital news publishers must capitalize on the traffic they do receive to keep their businesses afloat.  }

News publishers are increasingly reliant on digital advertising as print revenues decline~\cite{cheq2020}. However, even digital income is under strain: AI-driven search features have reduced referral traffic, with one 2025 report showing a 7\% drop in visits from Google year over year~\cite{dcn_ai_pubs}. This makes accurate brand safety classification vital, since mislabeling legitimate content as unsafe directly undermines publishers’ ability to monetize their limited traffic.


It is noteworthy that the aforementioned examples were the result of keyword-based brand safety tools. At the time, there was hope that newer AI and contextual-based brand safety tools would do more accurate classifications~\cite{cheq2020}. Today, AI-based brand safety classification is prevalent. Our paper investigates whether this technological advancement is satisfactory. 

The objective of this study is to analyze the effectiveness and consistency of the major brand safety providers: DV, IAS, and Oracle, to determine if they adequately protect the reputation of the advertisers' brand without losing advertising opportunities, and at the same time, prevent publishers' safe content from becoming demonetized. 



\begin{figure}[t!]
\centering
\includegraphics[width=1.0\linewidth]{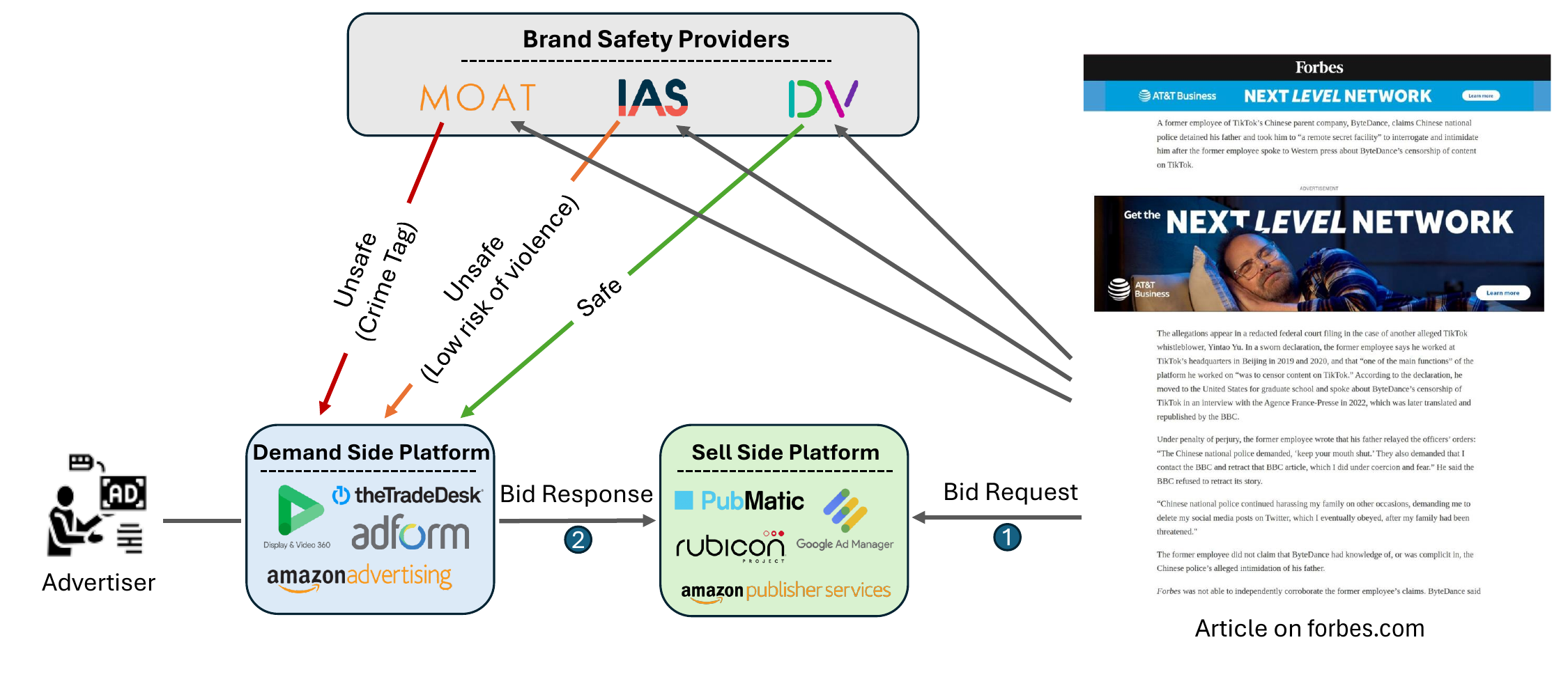}
\caption{\centering Three brand safety providers each classify an  article at Forbes.com}
\label{fig:brand_safety_classifiers}
\end{figure}

Our study shows that different providers may give different classifications for the same article. 
Figure \ref{fig:brand_safety_classifiers} shows how DV, IAS and Oracle categorize an article on Forbes' website.  
The article discusses crime, a topic that can be deemed unsafe.
While Oracle classifies the content as unsafe and related to crime, IAS deems it unsafe with a low risk of violence, and DV classifies the article as safe. This illustrates  the significant variation in the way brand safety tools interpret and classify the same content. If safety classifications are inconsistent, publishers would be confused about how to avoid demonetization, and advertisers do not know  what filters to set to avoid placement alongside undesirable content.



Few studies have evaluated the effectiveness of brand safety tools. One evaluated brand safety classifications on Reddit~\cite{vargo2024}, and another evaluated the brand safety classification of social media influencers' content~\cite{bishop2021}. Our research differs from these works in several aspects. First, instead of studying user-generated content on social media platforms, we study the brand safety classification of news articles on reputable websites. News content was selected because it is a major concern for advertisers, and digital advertising revenue is a key profit source for publishers. Since news content encompasses a broad range of topics and styles, our findings are generalizable. Second, we evaluate multiple third-party brand safety providers, which are widely used in the online advertising industry. Furthermore, we study many dimensions of brand safety (e.g., violence, crime, alcohol) instead of just considering general toxicity, as in prior literature.

\del{We choose to specifically study brand safety classification of news content because of digital advertising's growing importance to news publishers.}

Inspired by the Forbes Ad Inspector\footnote{https://forbes.github.io/ad-inspector/}, we built a data collection tool and collected brand safety data from 4,352 articles from popular news websites (e.g., nytimes.com, cnn.com, economist.com).

This study presents several major findings from our analysis of these 4,352 articles.
First, brand safety is a prevalent issue for the news media that the public may not be aware of. Our research indicates that brand safety providers have a significant impact on the advertising ecosystem and the news media publishers. Of the articles with brand safety data, 43.99\% were classified as ``unsafe'' by at least one of the three brand safety providers (Section~\ref{subsec:broad}). This impact is felt most by publishers that focus on breaking/general news, as we find that articles from these domains are more likely to be classified as unsafe (52.14\% of articles).

Second, we identified significant inconsistencies in the overall article classifications (i.e., whether the content is unsafe for any reason) among the brand safety providers. 
\del{Oracle tends to be more lenient, flagging fewer articles as unsafe (22.22\%) compared to DV (35.63\%) and IAS (53.72\%).} 
Disagreement in an article's brand safety rating occurs in 22.98\% of the cases where two providers rate the same article, and in 41.57\% of cases where all three providers give ratings. Furthermore, we evaluate interrater reliability with Krippendorff’s alpha, which adjusts for agreement that may occur by chance. The interrater reliability between all three providers, and each pair of providers, falls well below the conventional thresholds, indicating unreliable ratings\del{Notably, an article rated unsafe by a more lenient provider may be rated safe by a more stringent provider, or vice versa} (Section ~\ref{sec:overall_comparisons}). 

Third, an article may be unsafe for different categories, such as violence, adult content, hate speech, or drugs. 
Providers disagree not only on the assessment of an article's overall brand safety, but also on brand safety categories (see Table~\ref{tab:content_categories} for all categories). 
Different providers adopt different safety categories and different risk levels. \del{Oracle considers 10 categories, while DV considers 17, and IAS considers 27. The actual categories are related, but not the same.} Even when two providers consider the same safety category, their assessment of an article may be different.
Disagreements between two and three providers occurred in 62.40\% and 70.0\% of cases, respectively, when at least one provider rated the article as unsafe for the category.
Furthermore, the discrepancies vary significantly between different safety categories. 
%
%
The hate speech safety category, for example, was inconsistent 87\% of the time; while the death/injury category was inconsistent 57\% of the time (Section~\ref{sec:specific_comps_bw}).
%
%

Furthermore, our study shows that the current brand safety classification is not reliable even within the ratings assigned by the same provider.
For instance, IAS rates one article as unsafe regarding ``Illegal and Recreational Drugs'', but safe regarding ``Drugs'' (Section ~\ref{sec:specific_comps_wi}).

The absence of a common standard \del{exacerbates}may contribute to inconsistencies in brand safety classifications between providers. The Global Alliance for Responsible Media (GARM) previously provided a framework that defined a set of brand safety categories and risk levels \del{that should be considered when ensuring brand safety}to inform decision-making across platforms, advertisers, publishers, and brand safety providers. However, GARM was discontinued on August 9th, 2024 after a lawsuit from X (formerly Twitter)~\cite{garm-canceled}.
%
%

\del{In the absence of GARM, t}The Media Ratings Council (MRC) performs audits on several brand safety tools based on guidelines co-developed with the Interactive Advertising Bureau (IAB). However, these guidelines are not a comprehensive framework for brand safety and do not include specific definitions or examples of what constitutes safe or unsafe content~\cite{mrc-iab}. 
%
The need for a unified and transparent framework is underscored by the news in October 2024 that the US government is interested in investigating brand safety scandals \cite{usgov}, signaling broader concerns about systemic issues in the industry.



The contributions of our study are threefold. First, 
to the best of our knowledge, this is the first work that assesses widely used brand safety tools in news articles.
Second, our comprehensive analysis shows the prevalence of brand safety usage in the industry and, at the same time, the inconsistency\del{, inaccuracy, and unreliability} of the brand safety classification system for news articles. 
Furthermore, our findings call for action on standardizing guidelines in the space of brand safety. Specifically, a unified framework of brand safety categories between providers is needed. Reliable brand safety tools that are not self-contradicting are required, and greater consistency among different brand safety tools should be expected. 
A more standardized and transparent system, with explainable classifications, would reduce risks for advertisers, ensure fairness for publishers, and promote trust within the digital advertising ecosystem.

%% file: related.tex
\section{Related Work}
\label{sec:background_rw}



Brand safety has become a central concern for advertisers, as ensuring that ads do not appear next to harmful or inappropriate content is crucial to protecting a brand's reputation \cite{johnson2023,lee2021,marvin2018,griffin2023,hemmings2021}. Early approaches to brand safety relied largely on keyword-based filtering systems, which flagged content based on the presence of specific terms without accounting for context. Later, advanced methods that incorporate natural language processing (NLP) and machine learning (ML) were introduced to more effectively analyze the content \cite{vargo2024, griffin2023}.

Two works are related to ours in that they also evaluate brand safety tools, but in different contexts. The most relevant one examines Reddit’s brand safety classifications and compares them with toxicity scores calculated by the Perspective API\footnote{https://www.perspectiveapi.com/}. The authors find evidence of misclassifications, particularly on high-traffic forums, where potential ad revenues may influence safety classifications~\cite{vargo2024}. The second most relevant work studies influencer management tools, which help pair advertisers with social media influencers and rate the brand safety of social media influencers' content. This study found that one such tool used a list of keywords to calculate a ``family friendly rating'' and was systematically biased against LGBTQ creators due to words like ``queer’’ and ``gay’’ falling on the keyword list~\cite{bishop2021}.

Our study differs in several important aspects. We focus on ratings of news articles from multiple reputable websites (instead of user-generated content) and analyze how multiple brand safety providers classify the content (instead of a single provider). Unlike previous work, we examine classification variation across numerous brand safety dimensions (e.g., violence, crime, sexual content). 

In the larger literature on brand safety, several articles try to address how placement alongside unsafe content can affect advertising effectiveness or brand reputation. One paper finds that it does not decrease effectiveness in the case of advertisements shown before ``unsafe'' Youtube videos~\cite{bellman2018}. Others have found that the advertisement's effectiveness and brand's reputation are negatively affected when advertisements are placed next to offensive or unsafe content~\cite{johnson2023,lee2021}. These studies are orthogonal to ours, but show the importance of brand safety for advertisers.

Several works are devoted to the development of image-based classifiers to classify images and videos as unsafe with respect to adult content~\cite{vo2020,wang2018,jin2019,ou2017}, and additional sensitive safety categories such as gambling, guns, and gore~\cite{vo2020}. These works are also orthogonal to ours in that they attempt to build new classification systems for image and video content, while we audit existing systems that focus primarily on text content.

Our work relates to the broader literature on the shortcomings of algorithmic content moderation. Content moderation differs from brand safety classification in that it is applied to user-generated content instead of reputable news articles, and the goal is to protect users from toxic content (e.g., profanity or slurs) instead of advertisers from negative or inappropriate topics (e.g., violence). However, there are some similarities as the ultimate goal is to classify content based on its safety.
One study finds that automated content moderation tools struggle to replicate nuanced human moderation decisions in Reddit~\cite{Samory2021}. Another study finds that LLMs can improve content moderation performance over traditional NLP tools (e.g., Perspective API), but that performance varies significantly between different LLMs~\cite{kumar2024}. This finding echoes the inconsistencies we observe across brand safety providers.
Prior research has also found that misclassifications resulting from algorithmic content moderation are unequally distributed across demographic groups~\cite{gomez2024}. For example, the Perspective API tool was shown to unfairly penalize the content of LGBTQ influencers~\cite{oliva2021}. Similarly, unfair decisions against the LGBTQ, black, and blind communities have been found on platforms such as YouTube, TikTok, and Twitter~\cite{kingsley2022,ballburack2021, lyu2024, harris2023, vaccaro2021}.


Related literature has increasingly advocated for standardized, transparent, and contestable content moderation frameworks. Many researchers recognize that content moderation policy changes often happen silently, and advocate for greater transparency and clearer definitions of what is and is not allowed~\cite{pisharody2025, harris2023,gomez2024}. Several papers further propose explainable algorithms for content moderation so users get detailed answers for why their content was targeted~\cite{kingsley2022,gomez2024}.~\cite{vaccaro2021} propose mechanisms for making decisions contestable by affected users. Our paper supports this growing consensus and calls for standardization, transparency, explainability, and mechanisms that allow for appealing incorrect decisions in brand safety classification protocols \del{standardized content moderation protocols to improve fairness and accountability}.

%% file: method.tex
\section{Method}
\label{sec:method}

\begin{figure}[b!]
\centering
\includegraphics[width=1.0\linewidth]{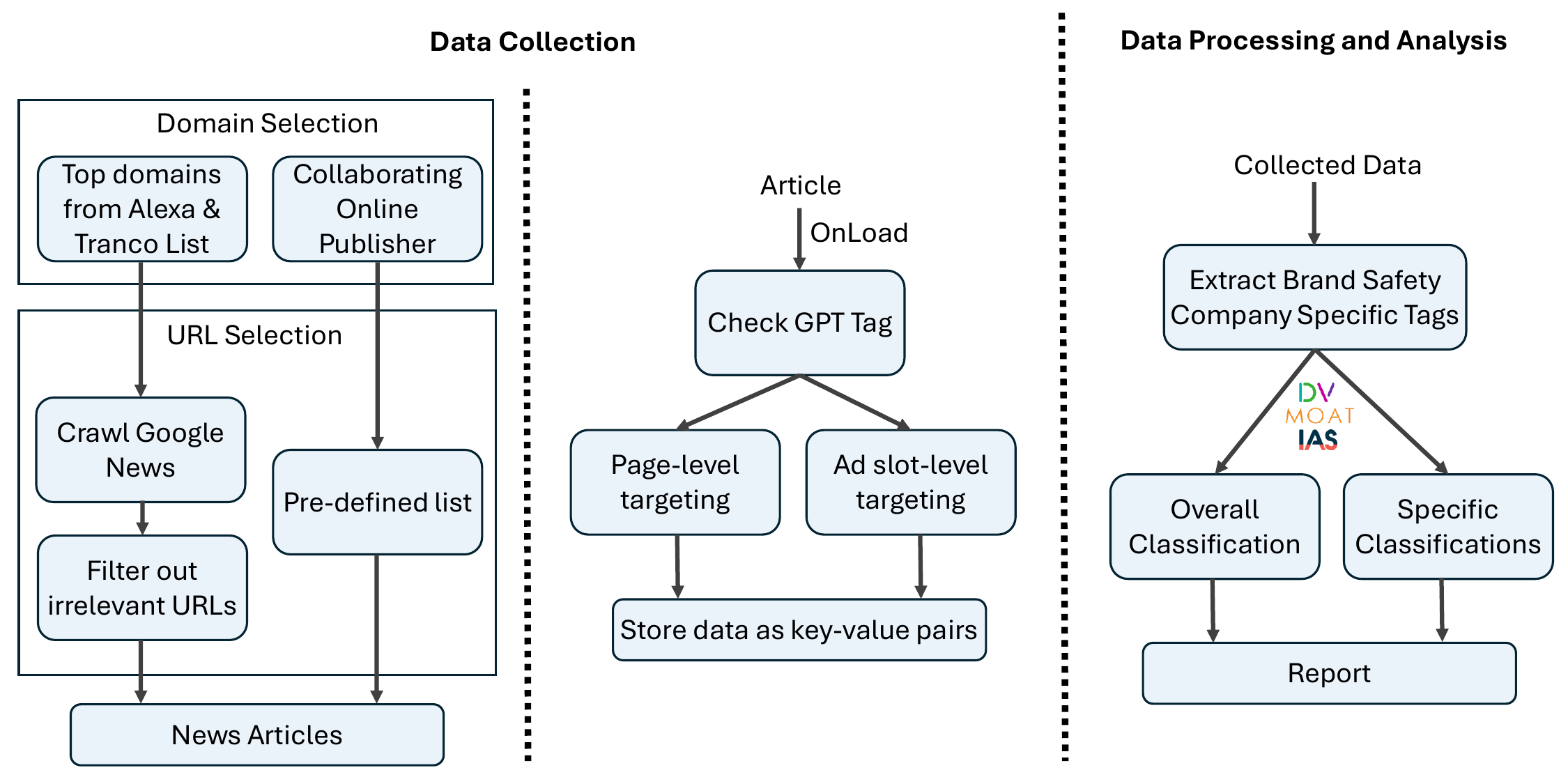}
\caption{System Design}
\label{fig:system_design}
\end{figure}

This section presents the methodology used to collect and analyze data for this study. First, Section~\ref{subsec:brand_safety_function} provides a brief overview of how brand safety tools work. Then, we describe our system, illustrated in Figure~\ref{fig:system_design}, in Section~\ref{subsec:data_collection} (\textit{data collection}) and Section~\ref{subsec:data_processing} (\textit{data processing}). Data collection involves two specialized web crawlers to collect brand safety data in popular online news domains. 
Data processing involves several data cleaning steps to extract provider-specific brand safety details, which are then used in the analysis presented in Section~\ref{sec:results}. The Github repository with the processed dataset and complete code for both crawlers will be made available after the review process.



\subsection{How Brand Safety Tools Work}
\label{subsec:brand_safety_function}

Brand safety providers collect page text and metadata (e.g., title) for each article, and use proprietary AI and natural language processing (NLP) models, as well as rule-based heuristics to assign risk levels across predefined safety categories. These classifications are stored  on the provider's servers and  are used to protect advertisers from placing their ads next to unsafe content.

There are two primary ways for advertisers to obtain ad inventory: through real-time bidding (RTB) auctions, or through direct deals with publishers (e.g., guaranteed deals). Publishers often manage their ad inventory using  Google Ad Manager (GAM), which helps publishers decide  whether to show an ad from an advertiser who has a guaranteed deal, or to show the winning ad from an RTB auction. In an RTB auction, safety classifications are shared with the demand side platform (DSP), who represents the advertiser. The DSP will not bid on behalf of an advertiser if the impression does not meet the advertiser’s predefined brand-safety requirements. 
\del{The brand safety data shared with the DSP server-side  is not visible to the webpage visitor. }
Because this exchange happens server-side, these classifications are not visible to external observers.

In cases where publishers integrate brand safety tools directly with GAM via Google Publisher Tags (GPT), the process occurs on the client side, allowing us to observe the classifications. In this setup, the publisher’s code issues an HTTP request to the brand safety provider when an article is loaded. The provider either retrieves a pre-computed rating from its database or performs the classification in real time and returns the result. These results are then appended to the HTTP request sent to GAM via the GPT tag, where they can influence the ad decision logic. The GPT integration thus enables both publishers (and researchers) to access page-level brand safety scores during article rendering. 

\del{Some websites integrate brand safety tools into their Google Ad Manager (GAM) via Google Publisher Tags (GPT). 
If advertisers with a guaranteed deal are concerned about brand safety, GAM can use brand safety classifications to prevent their ads from being shown on unsafe content. The integration of brand safety tools into GPT also allows publishers (and, by extension, our research team) to collect brand safety scores as the classifications are shown on the client side.}

\subsection{Data Collection Phase}
\label{subsec:data_collection}
The first phase of our system  focuses on 1) collecting the URLs of articles from various news domains, and 2) collecting the GPT data for the articles.

\subsubsection{Domain Selection}
We start with a list of 5,397 popular news domains developed by~\citet{usnewsdomains}, that was the union of Alexa's 1,000 top domains, domains frequently tweeted by US politicians, domains for local newspapers and television stations, and domains frequently visited by survey participants. This initial list does not rank the domains. 
We use the Tranco rankings~\cite{tranco}, which are widely adopted in academic research as an objective listing \del{ranking} of web domain popularity, to rank the initial list of news domains by popularity.

We then visit the top \del{first} 250 news domains \del{on the ranked news domains list}. We only visit the top 250, so our analysis uncovers brand safety data patterns across the most influential online news sources. 
For each domain, we visit it using \del{with}the Forbes Ad-Inspector tool, implemented\del{loaded} as a bookmarklet. This tool displays all of the Google Publisher Tag (GPT) data related to the ads on the page. We use the tool to determine whether brand safety data is available on each website.


If the publisher chooses to use the brand safety providers tags' with its Google Ad Manager ad server (see Section~\ref{subsec:brand_safety_function}), the brand safety data will be visible to us in the GPT data. Based on this filtering\del{In that case, we add the domain to a list of target domains. This results in a total of}, we identified 55 top news domains that we could capture brand safety data from. In addition, our collaborating online publisher (who requested to be anonymous) provides data for four of their domains. Thus, we have a total of 59 target domains for analysis. Notably, websites choosing not to collect brand safety data  does not affect whether advertisers can see this data. It only means that we can not access brand safety ratings on articles from these websites.


\subsubsection{URL Selection}
\label{sec:urls}
Next, we collect webpages of news articles from the 59 target domains. The collaborating publisher provided a list of around 150 URLs for each of their four
domains. For each of the remaining 55 target domains, we crawl 100 \textit{news articles} (i.e., not other content such as home pages or author pages) using Google News. Specifically, we \del{search} queried Google News \del{for articles in each such domain}with only the publisher’s domain name (e.g., cnn.com) and selected the first 100 articles returned.   
Note that Google News links are click-through URLs, meaning they redirect to the actual article pages. The crawler follows these links and extracts the article URLs.
For this purpose, we develop a JavaScript-based web crawler using the Playwright automation framework\footnote{https://github.com/microsoft/playwright} to crawl Google News and retrieve URLs for recent news articles on each target domain.
To refine our dataset, we excluded redirects to subscription pages (6 domains that always redirect to a subscription page were removed) and non-English websites (2 domains were removed).
Thus, 51 of the original 59 target domains remain, reflecting the filtering steps that were necessary to ensure the availability of brand safety data on news articles written in English.
With roughly 100 URLs for each of the 47 domains we crawled and 150 URLs for each of the four domains from our collaborating publisher, we have a total of 5,241 article URLs (not including those that redirect to subscription pages) from 51 domains. 

\begin{table*}[t]
\centering
\begin{tabularx}{\textwidth}{|>{\centering\arraybackslash}p{0.18\textwidth}|X|X|X|}
\hline
\textbf{Levels} & \centering\textbf{Oracle} & \centering\textbf{IAS} & \centering\textbf{DV} \tabularnewline \hline
Binary & Drugs, Death/Injury, Crime, Military, Adult, Arms, Terrorism, Hate speech, Tobacco, Obscenity & Politics (negative), Pop Culture (negative), Animal Cruelty, Discrimination Avoidance, Conspiracy Theories, Infectious Diseases, Protests, Misinformation, Pollution, Smoking, Natural Disasters & Any Moderate Content, Any Severe Content, Occult, Gambling, Pharmaceutical (negative), Financial (negative), Celebrity Gossip \tabularnewline \hline
Low, Medium, High & \centering - & Adult, Alcohol, Illegal Downloadable Material, Drugs, Hate speech, Offensive Language, Violence & Violence, Death/Injury, Drug Abuse, Crime, Adult/Sexual, Human–made Disaster, Terrorism, Hate speech/Cyberbullying, Alcohol, Natural Disasters, Vehicle Disasters, Profanity \tabularnewline \hline
\makecell[l]{Low, Medium, High, \\ Medium - News, \\ Medium - Entertainment, \\ Medium - Video Games} & \centering - & Death/Injury/Military, Crime, Sensitive Social Issues, Terrorism, Incitement of Hatred, Sexual Content, Arms/Ammunition, Cyber Security, Illegal/Recreational Drugs, Online Piracy, Obscenities & \centering - \tabularnewline \hline
\end{tabularx}
\caption{Brand Safety Categories for each Provider}
\label{tab:content_categories}
\end{table*}

\subsubsection{Collecting GPT data}
With the URL data set finalized, we proceed to the second phase, where we extract the \textit{advertising data} embedded on each article's web page. For this purpose, we develop a second crawler that visits each article's URL and calls on the Forbes Ad Inspector to collect the GPT data related to the ads on the web page. Brand safety data can be stored at different levels: ad-level or page-level. We collect brand safety data from both levels because different providers and publishers choose to store the data differently. The extracted data are stored locally for further analysis. Due to technical \del{challenges}constraints, such as metered paywalls, CAPTCHAs, broken links, and anti-crawling measures, we are unable to extract data from every article. \del{Ultimately}As a result, we successfully retrieve \del{the}GPT data from 4,352 of the 5,241 articles that we \del{visited}attempted to visit.





\subsection{Data Processing}
\label{subsec:data_processing}
After data collection, we process the data collected from 4,352 articles to extract brand safety data whenever available. Brand safety classifications are successfully retrieved\del{and have obtained it } from 3,219 articles. This subset of 3,219 articles from 51 top news domains, constitutes our final dataset for analysis.

All brand safety data are stored in key-value pairs. We show examples of the collected (Figure~\ref{fig:forbes_ad_inspector_data}) and processed (Figure~\ref{fig:brand_safety_data_samples}) data in Appendix~\ref{sec:images}. We collaborated with a Demand Side Platform (DSP) to get information on how brand safety data is stored and coded. This allows us to interpret brand safety data from the three brand safety providers: Oracle, IAS, and DV. 
Now we discuss the process of extracting brand safety data from each provider.


\subsubsection{Oracle}
\label{subsec:moat_categories_desc}

The extraction of Oracle brand safety data is the simplest. There is an overall binary rating for the article (that is, ``safe'' or ``unsafe'') stored under a key called \textit{ m\_safety}. Oracle also has a list of flags indicating which specific safety categories (e.g., crime) the article is unsafe for. This list of flags is stored under a key called \textit{ m\_categories}. The ten specific safety categories considered by Oracle are listed in the first row of Table~\ref{tab:content_categories} under the ``Oracle'' column. We extract the overall rating and the list of categories and store them in two separate variables. If the overall Oracle safety rating exists, we count the article as having Oracle brand safety data. In our data collection, there are 74 articles where the Oracle tag did not load properly and the value is ``waiting''.
We do not count these articles in our analysis, leaving us with 1,053 articles that have Oracle brand safety data.

\subsubsection{Integral Ad Science (IAS)}
\label{sec:ias_category_desc}
There are several components for collecting IAS brand safety data. In total there are 29 safety categories. The 11 safety categories in the first row of the ``IAS'' column in Table~\ref{tab:content_categories} are binary flags only indicating the presence of the category. These 11 categories are stored under the \textit{ias-kw} key. The other 18 safety categories can take on risk levels. The seven safety categories in the middle row of the ``IAS'' column in Table~\ref{tab:content_categories} are stored under separate keys. They can take on values of ``veryLow'', ``low'', ``medium'', or ``'high''. The ``veryLow'' value is the default value and means the article is safe with respect to the specific category, and the other values represent varying levels of risk with regard to the unsafeness of the article. We extract these seven safety categories and store each as their own variable. The remaining 11 safety categories (shown in the last row of the column ``IAS'' in Table~\ref{tab:content_categories}) are also stored under the \textit{ias-kw} key. These 11 safety categories have risk levels of low, medium, or high. However, they are different from the safety categories in the middle row of the ``IAS'' column in Table~\ref{tab:content_categories} as they have an additional classification that specifies whether the unsafe content is related to news, entertainment or video games.  

All brand safety values under the \textit{ias-kw} key are coded as a 7-digit number with `\_PG' appended. We decode these numbers using our DSP access. Some values stored under \textit{ias-kw} are just codes for keywords or contextual targeting categories that do not relate to brand safety, and thus, are removed. We store the list of decoded brand safety categories for each article in a variable called \textit{ias-kw-decoded}. 

IAS typically does not provide an overall safety rating for the article. Advertisers need to choose the safety categories they want to avoid (e.g. terrorism and politics), and check the safety rating with respect to those categories.
For the purpose of comparison with Oracle and DV, we construct an overall binary safety rating for each article: if the \textit{ias-kw-decoded} is empty and all seven of the separately stored category ratings take on a ``veryLow'' value, we consider the article to be rated safe by IAS, otherwise it is rated unsafe. 
For one domain (theglobeandmail.com), IAS instead stores values under the \textit{ias\_admants} key. For this publisher, IAS does not consider specific safety category ratings, but rates its overall safety  (through the presence/absence of a  ``brand\_unsafe'' string). 



As long as any of the IAS key-value pairs exist, we consider the article to have IAS brand safety data. This results in 1,666 articles with IAS brand safety data. For most articles, both the \textit{ias-kw} and seven variables in row two of Table~\ref{tab:content_categories} are scored. However, 80 articles only have \textit{ias-kw} data and 232 articles have the \textit{ias-kw} unscored. 

\subsubsection{DoubleVerify (DV)}
\label{sec:dv_categories_desc}
The DV brand safety data is  stored as a list of 8-digit codes under the \textit{BSC} or \textit{bsc} key. We use our DSP access to decode these 8-digit codes. We only retain the codes that are brand safety categories, and we store the list of decoded categories in the ``bsc-decoded'' variable. DV provides both specific category ratings and an overall article safety rating. If any brand safety category is present, there is an additional code for the ``Any moderate content'' category. There is also a code for ``Any severe content'', but we see this only twice in our dataset. If the ``Any moderate content'' code is present, we consider the article to be rated as unsafe. DV considers 17 safety categories. As shown in the first  row of the ``DV'' column in Table~\ref{tab:content_categories},
there are five categories that are only recorded for presence, and do not have an associated risk level. The second row lists the remaining 12 safety categories, where risk ratings are low, medium, or high. As long as the ``BSC'' or ``bsc'' key is not empty, we consider the article to have been rated by DV for brand safety. This results in 1,261 articles.

%% file: results.tex
\section{Results}
\label{sec:results}

Our data allows us to explore brand safety ratings from two perspectives: overall classifications and specific classifications by safety categories. Overall classifications categorize an article as ``safe'' or ``unsafe'' on a holistic level. Exploring brand safety at this level is important because some advertisers choose not to advertise on a web page where content is rated ``unsafe'' according to any category. However, it is also important to analyze brand safety ratings for specific safety categories (e.g., crime or hate speech)  because different brand safety providers consider different categories to define overall brand safety, and thus their overall brand safety ratings may not be directly comparable. Next, we present the results and insights for the overall and specific category classifications of the 3,219 articles obtained after processing the data as described in Section~\ref{subsec:data_processing}.

\subsection{Overall Classifications}
\label{subsec:broad}
 The analysis of brand safety ratings for the overall classification relies on the definitions of safe and unsafe that we derived for each provider, as discussed in Section~\ref{subsec:data_processing}. We find that 1,416 of the 3,219 articles in our sample (43.99\%) had an unsafe rating from at least one of the brand safety providers. This indicates that brand safety is having a large impact on the digital advertising industry. Our findings show that articles on domains publishing general/breaking news stories are more likely to be classified as unsafe compared with domains on specific news topics (52.14\% vs. 32.67\%). We also found that some brand safety providers are more stringent than others and, most importantly, that classifications are highly inconsistent between providers.

 \begin{figure}[b!] 
\centering
\includegraphics[width=1.0\linewidth]{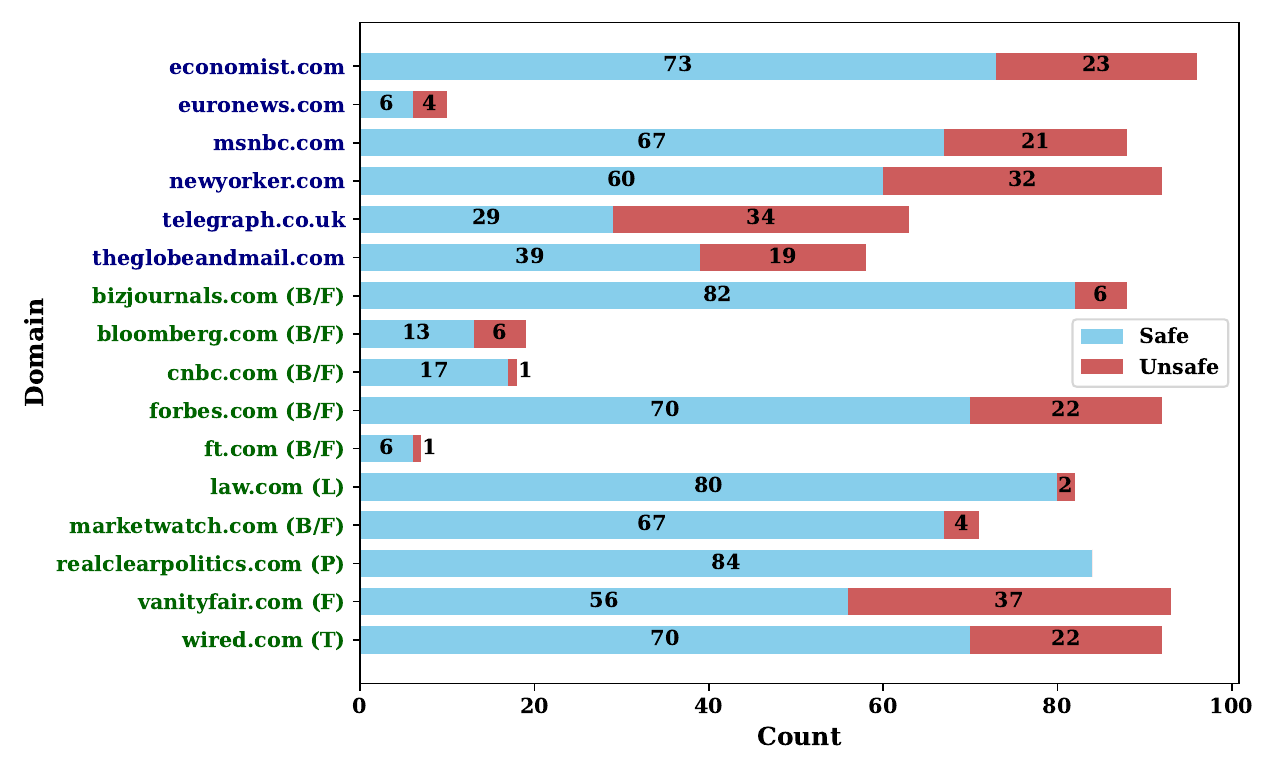}
\captionsetup{justification=centering}
\caption{Oracle ratings by domain\\ \footnotesize Note: Navy=Breaking/General News, Green=Specific Topic,\\
B/F=Business/Finance, L=Law, P=Politics, T=Technology, F=Fashion}
\label{fig:moat_ratings_domain}
\end{figure}
\subsubsection{Oracle}
The 1,053 articles with Oracle brand safety data are from 16 domains (see Section~\ref{subsec:moat_categories_desc}). Oracle rates  
234 (22.22\%) of these articles as unsafe. The breakdown of these safe and unsafe ratings by domain is shown in Figure~\ref{fig:moat_ratings_domain}. We categorize domains into different categories, such as general, breaking news, or a specific topic such as business, laws, politics, etc.
The details of labeling a domain into news categories can be found in Table~\ref{tab:website_categories} in Appendix~\ref{sec:appendix_desc}. 
There is a big difference in the percentage of articles rated as ``safe'' for domains that focus on a special topic (e.g., business/finance or law) versus general/breaking news. 
Articles on domains that focus on a specific topic are rated as safe by Oracle 84.37\% of the time. One domain (realclearpolitics.com) has all safe ratings, and another four domains, all focusing on a specific topic,  have over 90\% safe ratings. On the other hand, only 67.32\% of articles on domains in the general/breaking news category are rated as safe. This may be due to the fact that breaking news publishers are more likely to publish articles about inherently unsafe content (e.g., natural disasters, terrorist attacks, or wars). Of the 16 domains with ratings from Oracle, there is only one domain (telegraph.co.uk, a general news website) where safe ratings are in the minority (46.03\%).

\begin{figure}[b!]
\centering
\includegraphics[width=0.9\linewidth]{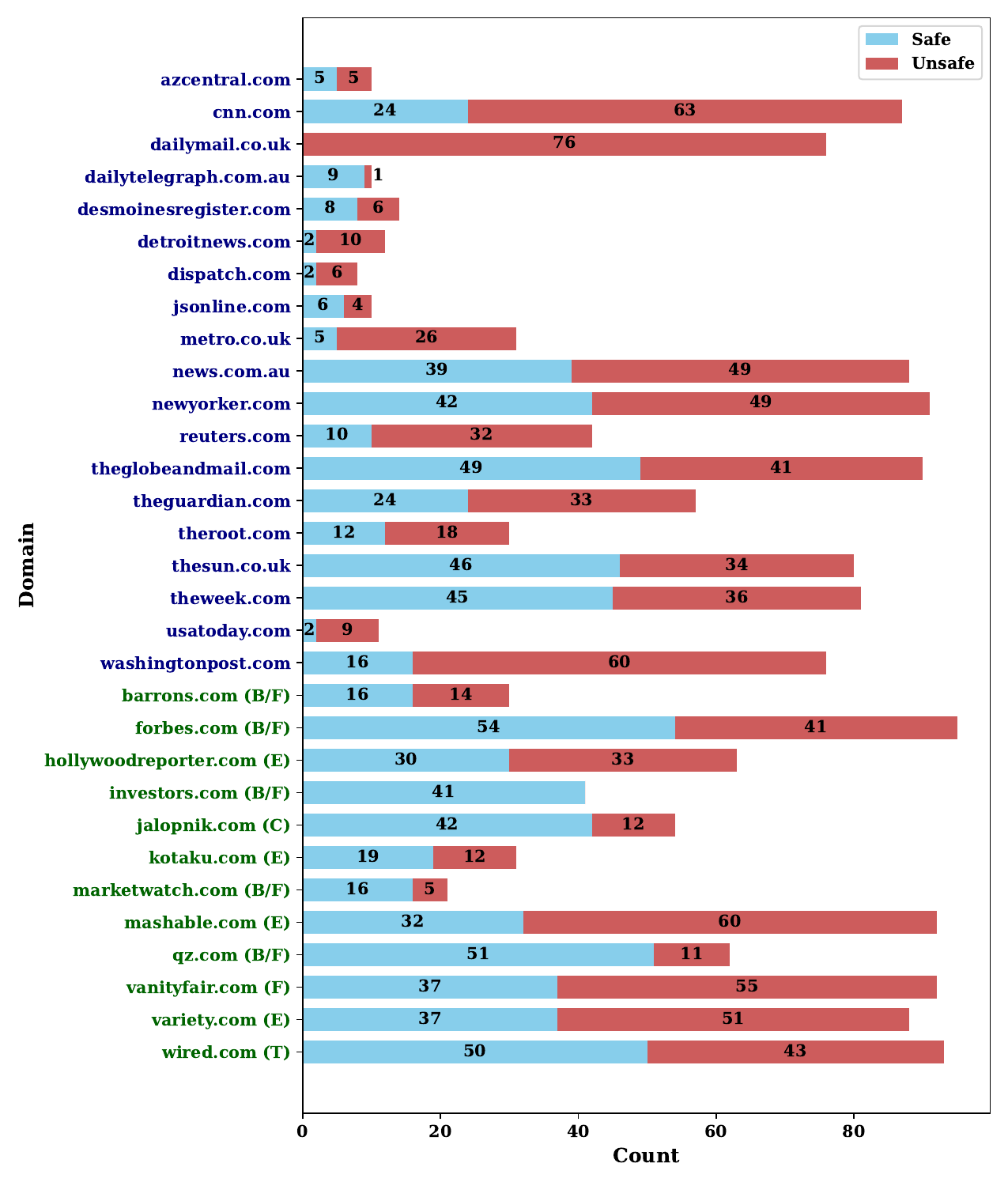}
\captionsetup{justification=centering}
\caption{IAS ratings by domain\\
\footnotesize Note: Navy = Breaking/General News, Green = Specific Topic,
B/F = Business/Finance, E = Entertainment, C = Cars, \\T = Technology, F = Fashion}
\label{fig:ias_ratings_domain}
\end{figure}

\subsubsection{Integral Ad Science (IAS)}

The 1,666 collected articles with IAS brand safety data are from 31 domains (see Section~\ref{sec:ias_category_desc}). IAS rates 895 (53.72\%) of these articles as unsafe. The breakdown of these safe and unsafe ratings by domain is shown in Figure~\ref{fig:ias_ratings_domain}. It is very different from Oracle's ratings (see Figure~\ref{fig:moat_ratings_domain}) in that there are 16 domains with more articles rated unsafe than safe by IAS. There is one domain with only safe ratings (investors.com), but there is also a domain with only unsafe ratings (dailymail.co.uk). The trend that articles from general/breaking news publishers are more likely to be rated unsafe still holds true for IAS. Articles on specific news topics domains are rated safe 55.77\% of the time, while articles on general/breaking news domains are rated safe only 38.27\% of the time. 


\subsubsection{DoubleVerify (DV)}

The 1,261 articles collected with DV brand safety data are from 16 domains (see Section~\ref{sec:dv_categories_desc}). 
DV rated 448 of these articles (35.27\%) as unsafe. There are two domains with over 90\% of the articles rated as safe,  both of which specialize in business/finance news. There are five domains where the majority of articles are rated unsafe,
and four of them are general news domains; the other is vulture.com, which focuses on entertainment. The domains that focus on a specific topic have articles classified as safe 77.9\% of the time, while the general/breaking news domains have articles classified as safe only 57.95\% of the time.

\begin{figure}[b!]
\centering
\includegraphics[width=1.0\linewidth]{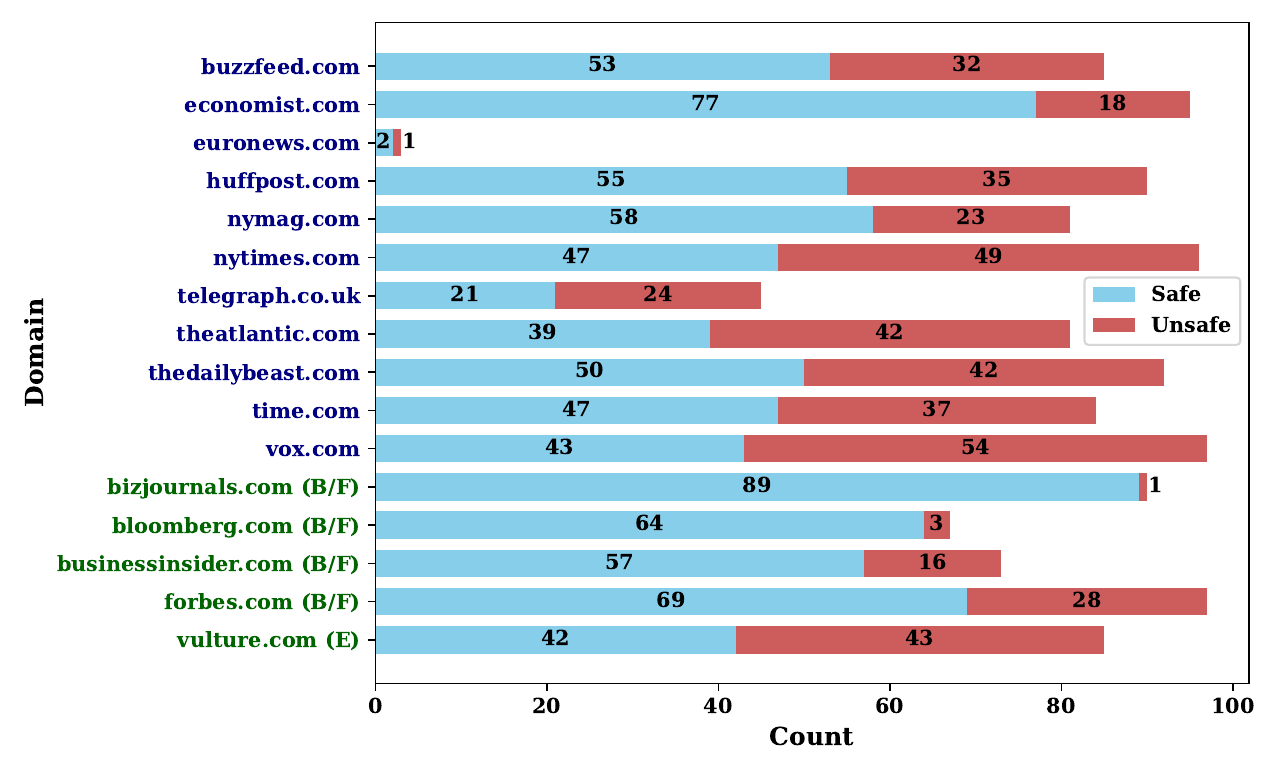}
\captionsetup{justification=centering}
\caption{DV ratings by domain \\ \footnotesize Note: Navy = Breaking/General News, Green = Specific Topic,
B/F = Business/Finance, E = Entertainment}
\label{fig:dv_ratings_domain}
\end{figure}

\subsubsection{Comparison between brand safety providers}
\label{sec:overall_comparisons}
We find that in general IAS is most stringent, with  the highest percentage of articles classified as unsafe, with DV second highest, and Oracle lowest. This may be due to IAS considering 29 brand safety categories, while DV and Oracle consider only 17 and 10, respectively (see Table~\ref{tab:content_categories}).

\begin{figure}[b!]
\centering
\includegraphics[width=\linewidth,trim=6.25cm 2.15cm 3.1cm 1.35cm,clip]{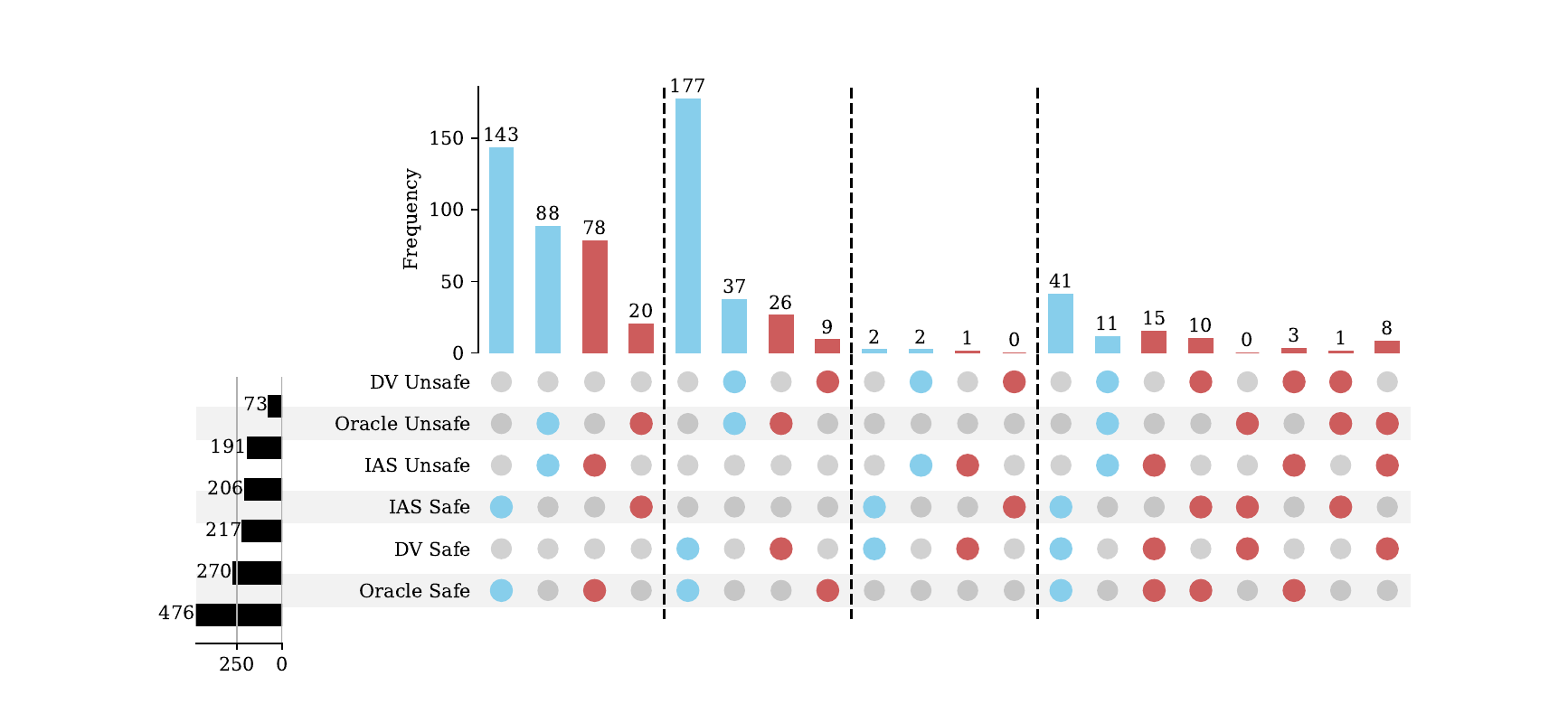}
\captionsetup{justification=centering}
\caption{Brand Safety Disagreements}
\label{fig:overall_disagreements}
\end{figure}

Next, we analyze the 672 articles that have multiple providers' brand safety ratings and compare different providers' ratings. There are 329 articles rated by just Oracle and IAS, 249 articles rated by just Oracle and DV, 5 articles rated by just IAS and DV, and 89 articles (all on forbes.com) rated by all three providers. In Figure~\ref{fig:overall_disagreements}, we use an UpSet style plot~\cite{lex2014upset} to show how each of these 672 articles is rated for overall safety by the different providers. The vertical dotted lines separate which providers are being compared in each section (e.g., the first section compares Oracle and IAS). Blue bars indicate agreement between providers, while red bars indicate disagreement. In total there is disagreement in 22.98\% of the 583 articles with ratings from only two providers. There is even higher disagreement when all three providers generate ratings: 41.57\% of the 89 articles.   

The first four bars in Figure~\ref{fig:overall_disagreements} break down (dis)agreements between Oracle and IAS on 329 articles rated by just these two providers. Disagreement is high (29.79\%). Such disagreements are not just due to different levels of tolerance for ``unsafe'' content. Across the entire dataset, IAS rates articles as unsafe at the highest rate, and Oracle at the lowest rate. However, there are 21 articles rated safe by IAS and unsafe by Oracle (including 1 article rated unsafe by all three providers). Furthermore, we  calculated Krippendorff's alpha ($\alpha$) to demonstrate the poor agreement between providers. Calculating interrater reliability is a more accurate measure of inconsistency than raw percentages because it accounts for agreement by chance. This is particularly important in our dataset where safe classifications are more likely than unsafe classifications. IAS and Oracle had an interrater reliability of only 0.42, far below the recommended threshold of 0.8, and even below the absolute minimum acceptable threshold of 0.67 for ratings that are only used for tentative conclusions~\cite{krippendorff}.

The second set of four bars in Figure~\ref{fig:overall_disagreements} break down the (dis)agreements on 249 articles rated by just Oracle and DV. Disagreement is lower (14.06\%), but this is partially due to both providers rating the majority of articles as safe. This indicates potential agreement by chance. The interrater reliability is higher than any other pairing of brand safety providers ($\alpha=0.53$), but is still far below the recommended and minimum thresholds of 0.8 and 0.67, respectively.

The third section of bars in Figure~\ref{fig:overall_disagreements} break down the (dis)agreements on the five articles rated by just IAS and DV. There are so few articles because IAS and DV are only both present on 94 forbes.com articles in our sample and MOAT also rates 89 of these articles (shown in the final section of bars in Figure~\ref{fig:overall_disagreements}). Across all 94 articles rated by IAS and DV, the disagreement is high (37.23\%) and the interrater reliability is very low ($\alpha=0.19$).

Classifications of the 89 articles rated by all three providers is shown in the final section of Figure~\ref{fig:overall_disagreements}. Disagreement between the three providers is high (41.57\%), and the interrater reliability is low ($\alpha=0.35$).

\subsection{Specific Classifications}
The results from Section~\ref{subsec:broad} show the inconsistency of brand safety tools in assessing the overall safety of an article. However, this inconsistency can be partially explained by variation in the number and nature of safety categories assessed by each provider. Additionally, some
advertisers may only block a few specific brand safety categories, and do not care about the overall safety rating. For example, a beer company may care specifically about not advertising on news articles about alcoholism, alcohol abuse, drug abuse, or drunk driving accidents, and set up brand safety filters accordingly. Thus, in this section, we examine the inconsistency in brand safety classifications with respect to a particular safety category. A descriptive analysis of which safety categories are most frequently occurring under each provider in our dataset is presented in Appendix~\ref{sec:appendix_freq}. 

\subsubsection{Comparison between brand safety providers}
\label{sec:specific_comps_bw}

Each provider considers different safety categories and different risk levels (see Table~\ref{tab:content_categories}). To compare the providers, we identify several brand safety categories that exist in the list of at least two brand safety providers (regardless of risk level). We make these pairings conservatively, and only pair categories from different providers when the category is not ambiguous. For example, we do \emph{not} pair the ``Drugs'' category for IAS with the ``Drug Abuse'' category from DV because drug abuse content is more severe than drug-related content. We do, however, allow for combining two categories under one brand safety provider to make a match. For example, IAS has a ``Death, Injury or Military Conflict'' category. While Oracle does not have this exact category, it does have ``Military'' and ``Death \& Injury'' categories. We combine these two categories for Oracle to allow comparison with IAS. The full list of pairings is displayed in Table~\ref{tab:categorical_comparisons}. For each safety category listed, we can tell which providers are compared from the ``Oracle'', ``IAS'', and ``DV'' columns. The ``T'' values indicate which provider is being compared for the specific category.  

\begin{table}[h]
\resizebox{1.0\linewidth}{!}{
\begin{tabular}{|>{\centering\arraybackslash}m{0.39\linewidth}|>{\centering\arraybackslash}m{0.115\linewidth}|>{\centering\arraybackslash}m{0.075\linewidth}|>{\centering\arraybackslash}m{0.06\linewidth}|>{\centering\arraybackslash}m{0.28\linewidth}|
>{\centering\arraybackslash}m{0.08\linewidth}|}
\hline
\textbf{Category} & 
\textbf{Oracle} & 
\textbf{IAS} & 
\textbf{DV} & 
\textbf{Inconsistencies}&
\textbf{$\alpha$}\\
\hline
Arms/Ammunitions & 
T & 
T & 
F & 
6/9 (66.67\%)&0.49 \\
\hline
Drugs & 
T & 
T & 
F & 
8/9 (88.9\%)& 0.19\\
\hline
Adult & 
T & 
T & 
F & 
14/23 (60.87\%)&0.54 \\
\hline
Adult/Sexual* & 
 F & 
T & 
T & 
6/9 (66.7\%)&0.47 \\
\hline
Natural Disasters & 
 F & 
T & 
T & 
1/1 (100\%)& 0.0\\
\hline
Terrorism & T & T & T & 2/3 (66.67\%)  & 0.66\\
\hline
Terrorism & T & T & F & 12/18 (66.67\%) &0.48 \\
\hline
Terrorism & T & F & T & 5/12 (41.67\%) & 0.73\\
\hline
Terrorism & F & T & T & 2/3 (66.67\%) & 0.49\\
\hline
Crime & T & T & T & 5/7 (71.43\%) & 0.56 \\
\hline
Crime & T & T & F & 37/61 (60.66\%)& 0.50 \\
\hline
Crime & T & F & T & 22/41 (53.66\%)&0.6  \\
\hline
Crime & F & T & T & 3/7 (42.86\%) &0.7 \\
\hline
Death/Injury & T & F & T & 24/42 (57.14\%) & 0.56\\
\hline
Death/Injury/Military* & T & T & F & 35/63 (55.56\%) & 0.55\\
\hline
Hate speech & T & T & F & 27/31 (87.10\%) & 0.19 \\
\hline
Violence & F & T & T & 19/26 (73.08\%) & 0.31 \\
\hline
Alcohol & F & T & T & 1/2 (50\%) & 0.66\\
\hline
Obscenities & T & T & F & 2/2 (100\%)&  0.00\\
\hline
\hline
\textbf{Total} & - & - & - & \textbf{231/369 (62.60\%)} &- \\
\hline
\textbf{-} & \textbf{T} & \textbf{T} & \textbf{F} & \textbf{141/216 (65.28\%)} & \textbf{0.37}\\
\hline
\textbf{-} & \textbf{T} & \textbf{F} & \textbf{T} & \textbf{51/95 (53.68\%)} & \textbf{0.63}\\
\hline
\textbf{-} & \textbf{F} & \textbf{T} & \textbf{T} & \textbf{32/48 (66.67\%)} & \textbf{0.44}\\
\hline
\textbf{-} & \textbf{T} & \textbf{T} & \textbf{T} & \textbf{7/10 (70\%)} &  \textbf{0.61}\\
\hline
\end{tabular}
}
\caption{\centering{Comparing Brand Safety Ratings for Specific Categories} }
\caption*{Note: * means two categories were combined}
\label{tab:categorical_comparisons}
\end{table}

After identifying the set of safety categories supported by all brand safety providers, we compare their respective classification outcomes for each shared category. For each category, we locate all articles with ratings from both providers and compute interrater reliability using Krippendorff's alpha (see ``$\alpha$'' in Table~\ref{tab:categorical_comparisons}). For interpretability, we also compute the inconsistency as a percentage. The \del{vast }majority of articles are not related to any given category and thus providers are in agreement by default. Therefore, we only consider articles with at least one unsafe rating when reporting disagreement as a percentage. 

Since there is no risk level provided for many safety categories, we consider two providers to agree when they both rate the article as unsafe for the target category, irrespective of risk level (e.g., if one provider rates unsafe with low risk and the other rates unsafe with high risk, we consider them \del{to agree}in agreement). Thus, we are making a conservative estimate of the inconsistency between brand safety tools. While this analysis does not always enable us to \del{say which provider is at fault}reveal the provider that misclassified the article, manual evaluation of some articles shows that all providers are inaccurate in some cases. 

\del{The number of inconsistencies in providers' brand safety ratings for specific safety categories are reported in Table~\ref{tab:categorical_comparisons}.} 
In total, there are 231 inconsistencies on 369 instances (62.60\%). There is an inconsistency on 7 of the 10 instances with a rating from all three providers. The average interrater reliability across all comparable categories for the three providers ($\alpha$=0.53) and each pair of providers is always below the minimum threshold of 0.67. The highest average agreement ($\alpha$=0.63) is between Oracle and DV, which can be compared for three categories. The lowest average interrater reliability ($\alpha$=0.37) is between IAS and Oracle, which can be compared across eight categories. 

Digging in deeper, Table~\ref{tab:categorical_comparisons} shows 141 disagreements between IAS and Oracle, 65.28\% of all 216 instances where IAS and Oracle can be compared. \del{Of the 140 disagreements, t}There were 75 times where only Oracle flagged the category as unsafe, and  66 times only IAS flagged a category unsafe, with 39 times being low risk, 26 times being medium risk, and once being high risk. 
In the most egregious example, where IAS rated an article as high risk for drug content while Oracle rated it as 
safe, it appears IAS is at fault because the article covers an update to the Samsung Galaxy phones. Our manual inspection of this article revealed no reason for this. As an example where Oracle is at fault, an article in \textit{The New Yorker} about ``Black Identity'' is labeled by Oracle as being unsafe and related to crime. 

Table~\ref{tab:categorical_comparisons} shows  51 instances of an inconsistency between Oracle and DV ratings,  53.68\% of the 95 instances that Oracle and DV can be compared. Most disagreements are due to \del{the fact }Oracle \del{is}being more stringent, flagging a category as unsafe when DV rates it as safe (41 times). As an example, an article about the Hezbollah pager explosions cites 9 dead and 2,750 injured, yet DV does not rate the article unsafe for ``Death \& Injury''.

Finally, there are 32 cases of disagreement between IAS and DV, 66.67\% of the 48 instances that can be compared. IAS is a more stringent provider, flagging 26 out of the 32 instances \del{,}as unsafe\del{,} with a risk level of low 13 different times, medium 11 times, and high twice. For the other 6 disagreements (where DV rates the article as unsafe and IAS does not), DV rates the risk level as low once and medium five times.  

We can also look at the two safety categories (Terrorism and Crime) in Table~\ref{tab:categorical_comparisons} where all three brand safety providers can be compared. For ``Terrorism'', there are only three articles for comparison and two are disagreements. For ``Crime'', there are seven articles for comparison, and five have a disagreement. In three of the disagreements only Oracle rates the content as unsafe.

\subsubsection{Comparison among safety categories}
\label{sec:specific_comps_wi}
We further analyze the inconsistencies across different safety categories, as shown in Table~\ref{tab:categorical_comparisons}.
\del{Some categories like Drugs (87.5\%) and Hate Speech (87.1\%) have a very high percentage of inconsistencies. Other categories, such as Death/Injury (57.14\%) have a much lower percentage of inconsistencies.} The interrater reliability for all specific categories falls below the recommended reliability threshold of 0.8, but there are two instances (IAS and DV rating crime, and Oracle and DV rating terrorism) where the interrater reliability exceeds the minimum acceptable threshold of 0.67. The remaining 17 categories have interrater reliability scores that signal unreliable ratings, which\del{A higher rate of inconsistency} may indicate that the corresponding safety category is harder to classify. Under this assumption,  our results illustrate  which safety categories need more concrete definitions in an updated framework of brand safety. Brand safety providers should also \del{emphasize}prioritize improving classification accuracy within the\del{se} safety categories with a higher level of inconsistency.
\subsubsection{Comparisons within brand safety providers}
We also examine the consistency of behavior within a brand safety provider. 
For IAS, sometimes the category ratings are in conflict with each other. For example, there are separate ``Adult'' and ``Sexual Content''safety categories, and we expect these two categories to have  similar ratings. There are 179 articles where both categories are considered and at least one of them has a low, medium, or high risk rating. Of those 179 articles, there are:
\begin{itemize}
    \item 12 articles where both categories are rated unsafe with the same risk level.
    \item 12 articles where both categories are rated as unsafe, but with different risk levels.
    \item 155 articles where one category is safe and one is unsafe.
\end{itemize}
We consider some of these inconsistencies to be extreme. For example there are 36 articles where one category is rated as safe and the other is unsafe with a risk level higher than low (35 with medium risk and 1 with high risk). Looking into these 36 articles, we find that 12 of them discuss sexual assault, sex trafficking, or rape, and yet have a safe rating for either ``Adult'' or ``Sexual content''. In contrast, there is also an article about the construction plans for a highway in Arizona classified as medium risk for adult content.

There is a similar trend occurring for the ``Drugs'' and ``Illegal and Recreational Drugs'' categories. There are 46 articles where both safety categories are considered and at least one of the two drug-related categories has a low, medium, or high risk rating. There are only two articles where both categories have an unsafe rating, and the risk levels for the two categories are not the same in either case. 
For the other 44 articles, there are 33 times where ``Illegal and Recreational Drugs'' is rated as safe and ``Drugs'' is not (30 low risk, 2 medium risk, 1 high risk). There are 11 articles where ``Drugs'' is rated as safe and ``Illegal and Recreational Drugs'' is not (all medium risk). Looking at the 14 articles with medium or high risk we again find that some of these articles are not ambiguous. Three of the articles mention hard drugs such as ketamine or ecstasy, and one of the articles is the aforementioned article about Samsung Galaxy phones that is rated high risk for drug content. 

For DV, the aggregate category for severe content is present each time that there is high risk for alcohol or profanity (one of their occurrences is on the same article). However, DV does not classify the seven articles with high risk for ``Adult \& Sexual'' content as ``severe content''. It is unclear why being rated as unsafe with high risk for some categories and not for others would classify an article as ``severe content''. 

%% file: discussion.tex
\section{Discussion}
\label{discussion}

{\bf Issues in Current Digital Advertising Ecosystem.}
Each provider approaches brand safety with a different proprietary algorithm and set of brand safety categories. Combined with a lack of standardization and regulation, inconsistencies occur as expected. These problems are not limited to breaking news media. Our data show that inconsistency extends across many different categories such as entertainment, health, and finance. Inconsistent ratings are especially harmful to publishers as they can not reliably figure out how to prevent unsafe classifications, nor the resulting decline in advertising revenue. Advertisers can mitigate some of the harms of inconsistent ratings by relying on a single provider (publishers do not have this choice); however, we also find inconsistency within the ratings of a single provider.

We also manually identified instances of clearly incorrect brand safety ratings. These misclassifications carry serious consequences: advertisers risk damaging their brand when their ads are served next to harmful content misclassified as safe, and publishers lose revenue when valuable, contextually appropriate content is mislabeled as unsafe.
The consequences of misclassification in content moderation are complex, and prior research has shown that such misclassifications typically harm marginalized groups disproportionately~\cite{kingsley2022,oliva2021,gomez2024,harris2023}. This makes the fact that an article about ``Black Identity'' was labeled as unsafe and related to crime in our dataset unsettling. Although we did not have enough relevant articles to draw concrete conclusions, future research should investigate this potential problem more thoroughly.

{\bf Call for Industry Standardization and Academic Research.}
While our findings point to serious flaws in the current digital advertising landscape, 
advertisers still value safe\del{t}\rev{l}y-aligned ad\del{s} placements, and publishers need monetization tools that protect brand integrity without sacrificing legitimate content. 

Our findings call for a standardized, transparent framework for brand safety classification, which is essential for ensuring consistency, accuracy, and trust for a sustainable digital advertising ecosystem on the open web. This framework should include:

\begin{itemize}
    \item Clear definitions of safety categories, with concrete examples of what constitutes risks at different levels, 
    \item Auditable classification processes, where providers agree to regular third-party evaluations in exchange for industry accreditation, 
    \item An appeal mechanism for publishers to contest misclassification\rev{s},
    \item A public benchmark dataset, grounded in human-labeled brand safety ratings, to evaluate accuracy and bias across tools,
    \item Explainable classifications (i.e., a rationale for why content was rated safe/unsafe should be provided). 
\end{itemize}


While brand safety providers may initially resist such measures,  advertisers and publishers, who ultimately fund these tools, hold the leverage. If they collectively demand compliance with a standardized framework, non-compliant brand safety providers may be pushed out of the market. Importantly, such a system would still allow differentiation: providers could offer better classification quality and additional features such as specialized categories, fine-grained risk levels, or dynamic pricing models.


Our study advances academic and technical discourse on content classification and calls for deeper engagement from the Web, natural language processing, and machine learning communities to build culturally sensitive, explainable brand safety models, an urgent need amid increasing regulatory scrutiny~\cite{usgov}.


{\bf Limitations of the Study.}
While our study presents  valuable findings of inconsistencies in brand safety classification among leading providers, several methodological and conceptual limitations should be noted.


Although our analysis is limited to English-language news articles, the brand safety systems we evaluate are not confined to this domain. These tools are widely deployed across international markets and are applied to various formats, including video, podcasts, and user-generated content. Consequently, the methodological concerns we identify—classification inconsistency and the absence of ground-truth benchmarking—are likely to extend beyond our dataset and may affect brand safety assessments across languages, formats, and cultural contexts. Investigating how classification systems perform across these diverse types of media would provide a more comprehensive understanding of brand safety practices in the digital ecosystem.

Second, our primary analytical focus is on classification inconsistencies between brand safety providers rather than classification accuracy. This is due to the absence of a ground-truth dataset of brand safety labels against which we could benchmark provider's output. Without independent, human-verified assessments of content safety, we cannot definitively determine whether any given provider's classification is correct or incorrect. We call for future research to develop such a benchmark using human expert annotation. Developing a human-annotated benchmark to evaluate the accuracy of brand safety classifications remains an important avenue for future work, and we are actively exploring best practices for designing such a dataset in consultation with industry providers. Recent work indicates that carefully calibrated moderation interventions can steer users away from extremist movements~\cite{Russo2025}, highlighting the broader social implications of reliable classification of safe content.

Third, our dataset only includes webpages for which their publishers choose to collect brand safety data from the providers. 
This introduces a potential selection bias.
Consequently, our findings on the proportion of articles labeled unsafe may be influenced by publishers’ self-selection, potentially skewing our understanding of the overall coverage of brand safety. Our analysis assumes that the publisher's decision to collect brand safety data is independent of the safety classification provided by a brand safety provider, a conceptual constraint that we acknowledge and plan to test in future research.

%% file: conclusion.tex
\section{Conclusions and Future Work}
\label{conclusion}

This paper presented an empirical analysis of brand safety classifications from three industry leading providers, DoubleVerify (DV), Integral Ad Science (IAS), and Oracle. Our findings show that brand safety has a large effect on the digital advertising industry as there was an unsafe rating from at least one provider on 43.99\% of the articles in our dataset (including 52.14\% of general/breaking news articles). The frequency of unsafe ratings is not a problem in itself, but the lack of standardization and consistency in the ratings is. We found inconsistent classifications between two providers in both overall safety assessments (22.98\%) and individual risk categories (62.40\% of the time where at least one unsafe rating is present). Low interrater reliabilities, as measured by Krippendorff’s alpha, indicate that both the overall safety, as well as the specific category-level, ratings from brand safety providers are unreliable. These inconsistencies are not isolated anomalies, but indicate deeper issues stemming from inherent technical challenges, proprietary algorithms and varying interpretations of what constitutes ``unsafe'' content. \del{Finally, we even showed that IAS often produces ratings that conflict with conflicting classifications (e.g., 155 articles where one of ``Adult'' and ``Sexual Content'' categories is rated as safe, and one is rated as unsafe).} Finally, we showed that IAS sometimes produces internally inconsistent ratings. For instance, there are 46 articles where IAS rated as safe for the ‘Drugs’ category but as unsafe for the ‘Illegal and Recreational Drugs’ category, or vice versa.

Such inconsistencies carry serious consequences and undermine the trust in brand safety technologies that are central to the operation of the digital advertising ecosystem. If publishers and advertisers cannot correctly identify unsafe content,  
publishers cannot mitigate the revenue losses associated with  content that is classified as unsafe, and advertisers may reduce their digital advertising budget to protect their brand's reputation. Furthermore, some inconsistencies are due to blatant misclassifications, which directly harm publishers' revenues when safe content is incorrectly flagged as unsafe, and advertisers' reputations when their ads are shown next to unsafe content that is misclassified.

Although our dataset is primarily made up of English-language web articles,
the issues that we have identified of inconsistent classification of safe content both within and between providers are fundamental to  inherent technical challenges in classification and lack of standardization. Therefore, we believe that our results are broadly generalizable across different content, geographies and media types.

To address these issues, we advocate for the implementation of a standardized, transparent, explainable, and auditable brand safety framework. Such a framework would include clearly defined safety categories, routine third-party audits, explainable classifications, and an appeal process for misclassified content.
By adopting a shared protocol, the industry can move toward more reliable and fair evaluations that serve the interests of advertisers and publishers.

Our findings also raise important areas for future research. Content safety classification poses inherent technical challenges to natural language understanding. The necessity of evaluating words in their context (e.g., ``shoot'' has different meanings in a sports article versus a breaking news story about a violent criminal) lends itself well to large language models. However, open research questions remain about how to effectively handle hierarchical classification (e.g., assigning low, medium, or high risk levels to certain categories), mitigate potential biases against specific dialects or writing styles, and generate explanations for each classification that are both faithful and plausible.


%% file: acknowledgements.tex
\section*{Acknowledgements}
This work was supported by the NSF under Grants No. CNS 2237328 and DGE 2043104, as well as by the Martin Tuchman’62 Chair Endowment and the Leir Foundation.  ChatGPT and Writefull were utilized to refine some sentences for clarity, grammar, spelling, and punctuation.

%% file: appendix.tex
\appendix

\section*{Appendix}
\label{appendix}
\section{Examples of Collected and Processed Data}
\label{sec:images}
Figures 6 and 7 provide illustrative examples of the brand safety datasets referenced in our study. Figure 6 displays the original GPT-generated classifications for IAS, while Figure 7 shows the corresponding processed format prepared for analysis.
\begin{figure}[H]
\centering
\includegraphics[width=0.5\linewidth]{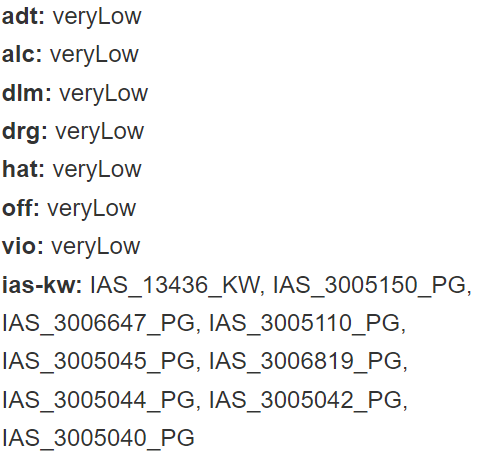}
\caption{\centering{Example of Collected GPT Brand Safety Data for IAS}}
\label{fig:forbes_ad_inspector_data}
\end{figure}

\begin{figure*}[h]
\centering
\includegraphics[width=1.0\linewidth]{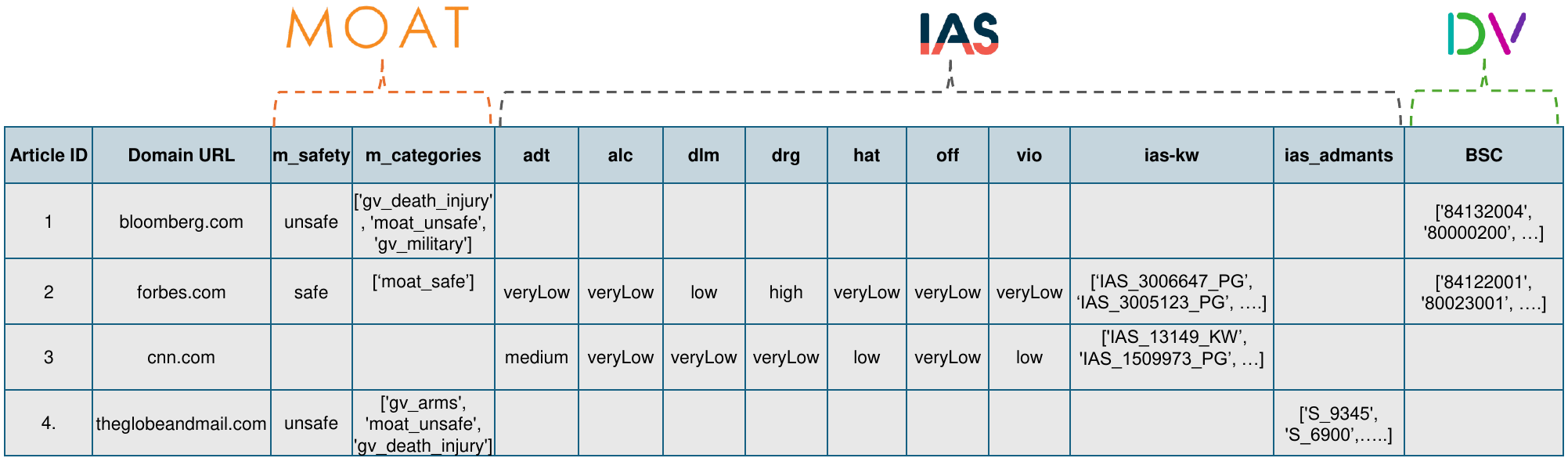}
\caption{Example of Processed Brand Safety Data}
\label{fig:brand_safety_data_samples}
\end{figure*}

\section{Frequency of Specific Brand Safety Categories}
\label{sec:appendix_freq}

Given we could not find any previous works that collected brand safety data for news articles, we feel there is some benefit in presenting descriptive results. For example, we showed that articles on breaking/general news websites are more likely to be rated as unsafe than articles from domains which focus on a specific topic (e.g., business). For this same reason, we want to present the specific content categories which are most likely to be rated as unsafe on news articles. These results are not important to our main analysis, thus we present them here in the Appendix. 

\subsection{Oracle}
\begin{figure}[H]
\centering
\includegraphics[width=1.0\linewidth]{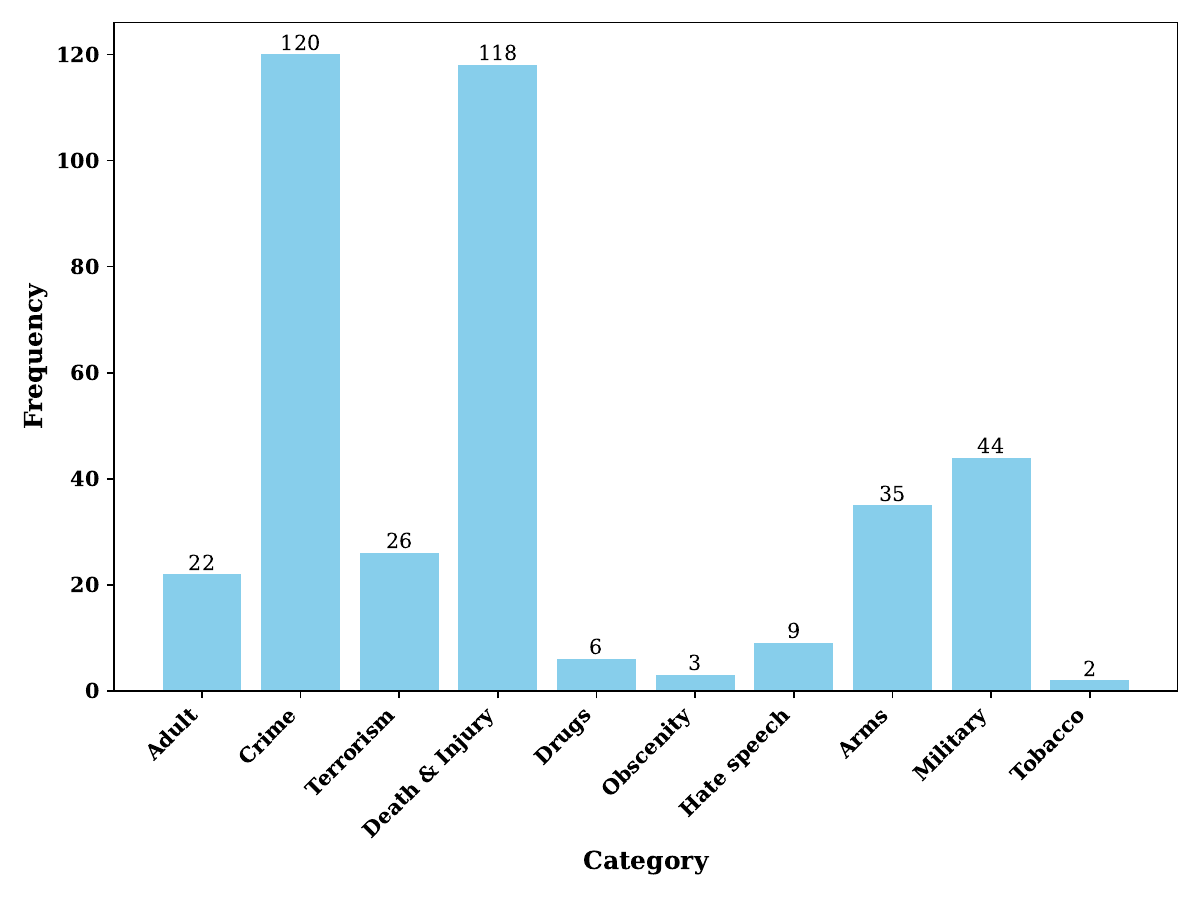}
\caption{Oracle Specific Classifications}
\label{fig:moat_categories_appendix}
\end{figure}

Oracle has the simplest specific classifications system of the three providers. There are only 10 categories, as shown in Table ~\ref{tab:content_categories}. Classification into each category is only binary, and there are no risk levels. The frequency with which each of these 10 categories is found in the 234 articles classified as unsafe by Oracle are visualized in Figure ~\ref{fig:moat_categories_appendix}. The ``Crime'' and ``Death \& Injury'' columns are by far the most common, with 120 and 118 occurrences, respectively. 

\begin{figure}[H]
\centering
\includegraphics[width=1.0\linewidth]{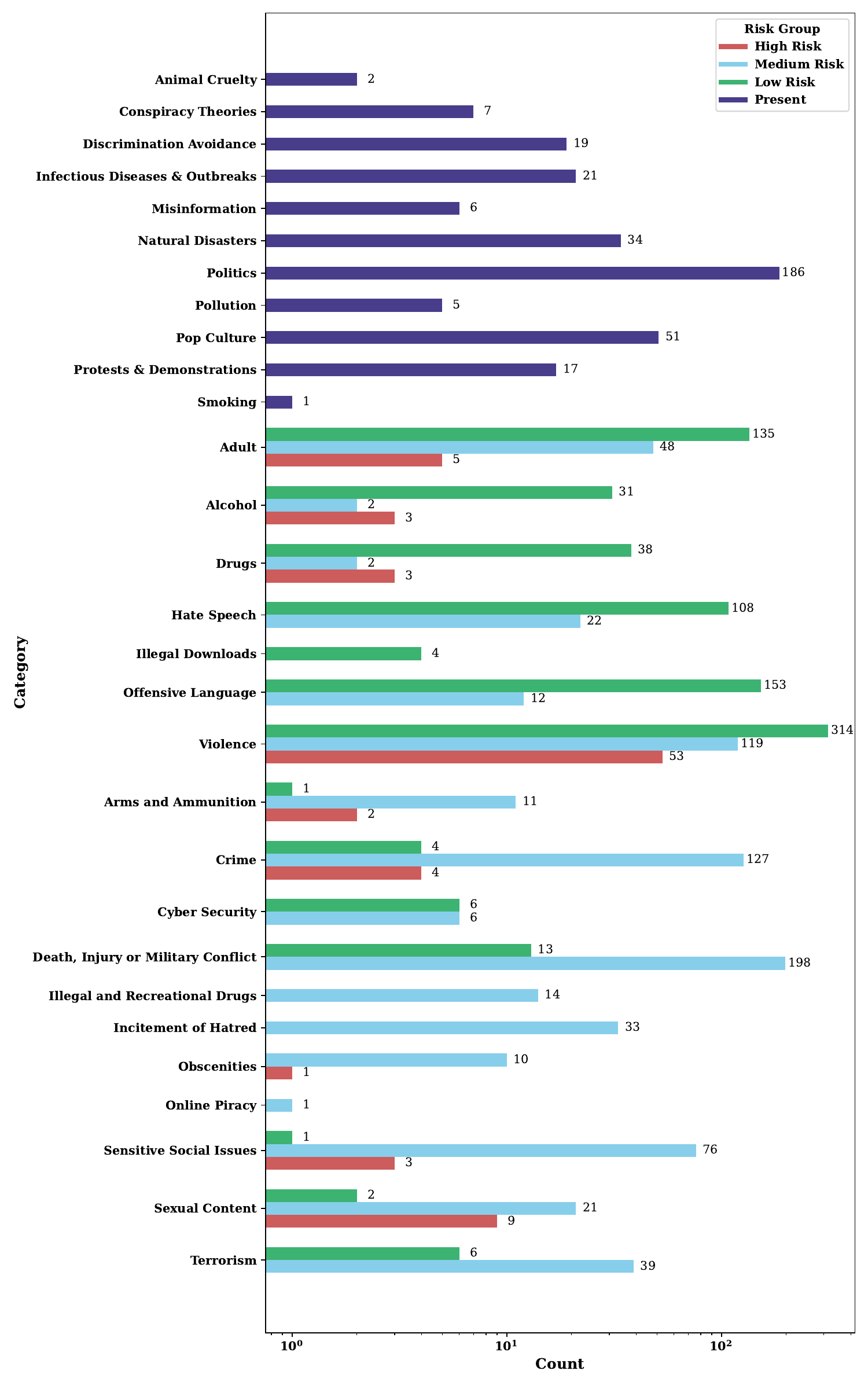}
\caption{IAS Specific Classifications}
\label{fig:ias_categories_appendix}
\end{figure}

\subsection{Integral Ad Science (IAS)}
\begin{figure}[h]
\centering
\includegraphics[width=1.0\linewidth]{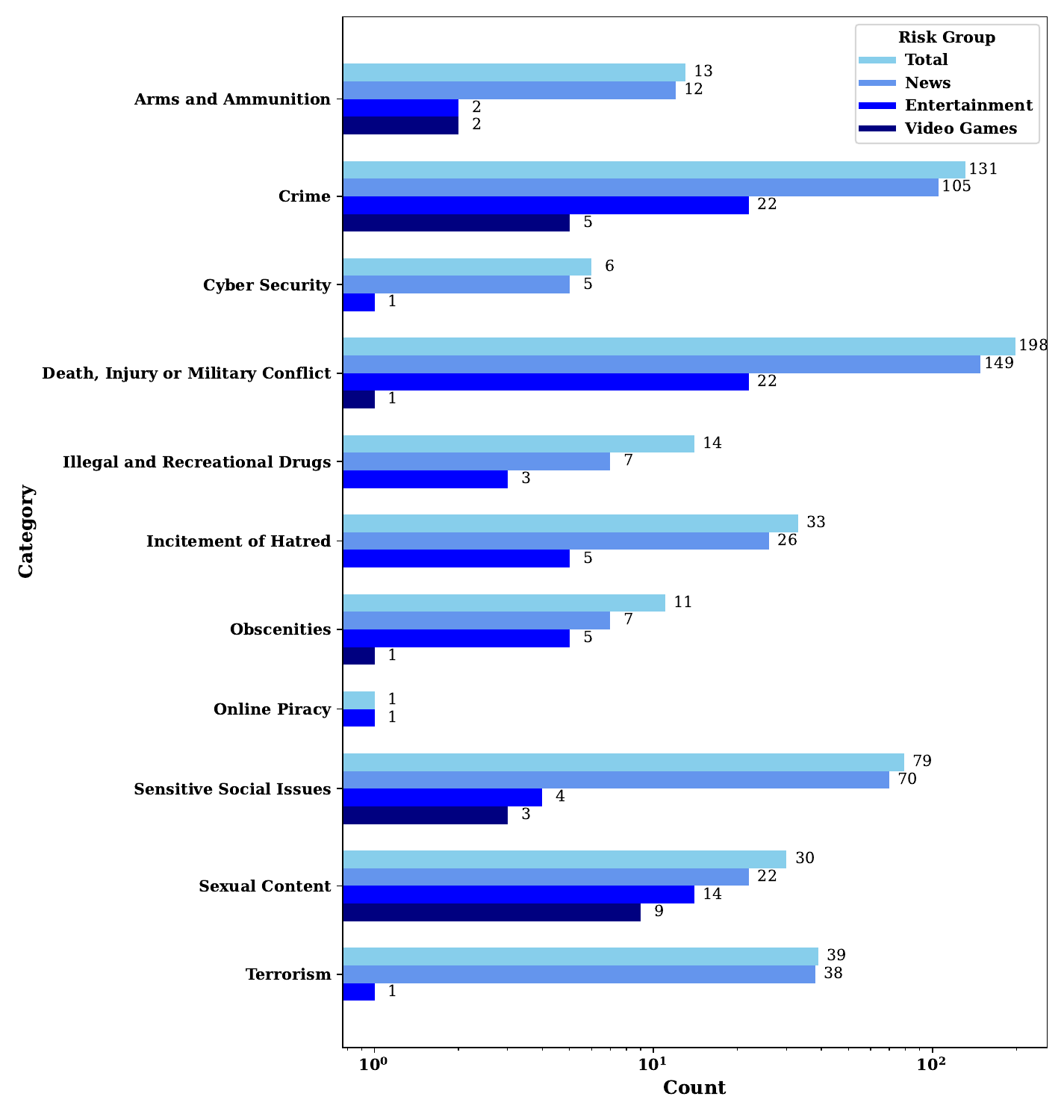}
\caption{\centering{IAS Specific Classifications with News, Entertainment, and Video Gaming Tags}}
\label{fig:appendix_ias_tags}
\end{figure}

The IAS specific content category classification system is the most complex of the three providers as described in Section ~\ref{sec:ias_category_desc}. We show the frequency of all 29 IAS categories in the 895 articles rated as unsafe by IAS in Figure ~\ref{fig:ias_categories_appendix}. Note that we do not consider the additional ``News'', ``Entertainment'', or ``Video Gaming'' tags and just plot each category according to risk level. We do separately show the frequency of the additional ``News'', ``Entertainment'', or ``Video Gaming'' tags in Figure ~\ref{fig:appendix_ias_tags}.

Of the binary categories which are only measured for presence, Figure ~\ref{fig:ias_categories_appendix} shows that ``Politics'' is by far the most common with 186 occurrences. For the categories allowing a risk level to be assigned, categories with over 100 total classifications include ``Adult'', ``Hate speech'', ``Offensive Language'', ``Violence'', ``Crime'', and ``Death, Injury, or Military Conflict''. There are some similarities to Oracle, where ``Crime'' and ``Death \& Injury'' were the two most common categories. However, ``Adult'' and ``Hate Speech'' were not very common in the Oracle classifications. Of the 18 categories with a risk level assigned, we find that ``High Risk'' is rarely used. The one exception is that 53 articles are rated as high risk with respect to the ``Violence'' category. There are only 30 high risk ratings among the other 17 categories combined.

\subsection{DoubleVerify (DV)}

\begin{figure}[h]
\centering
\includegraphics[width=1.0\linewidth]{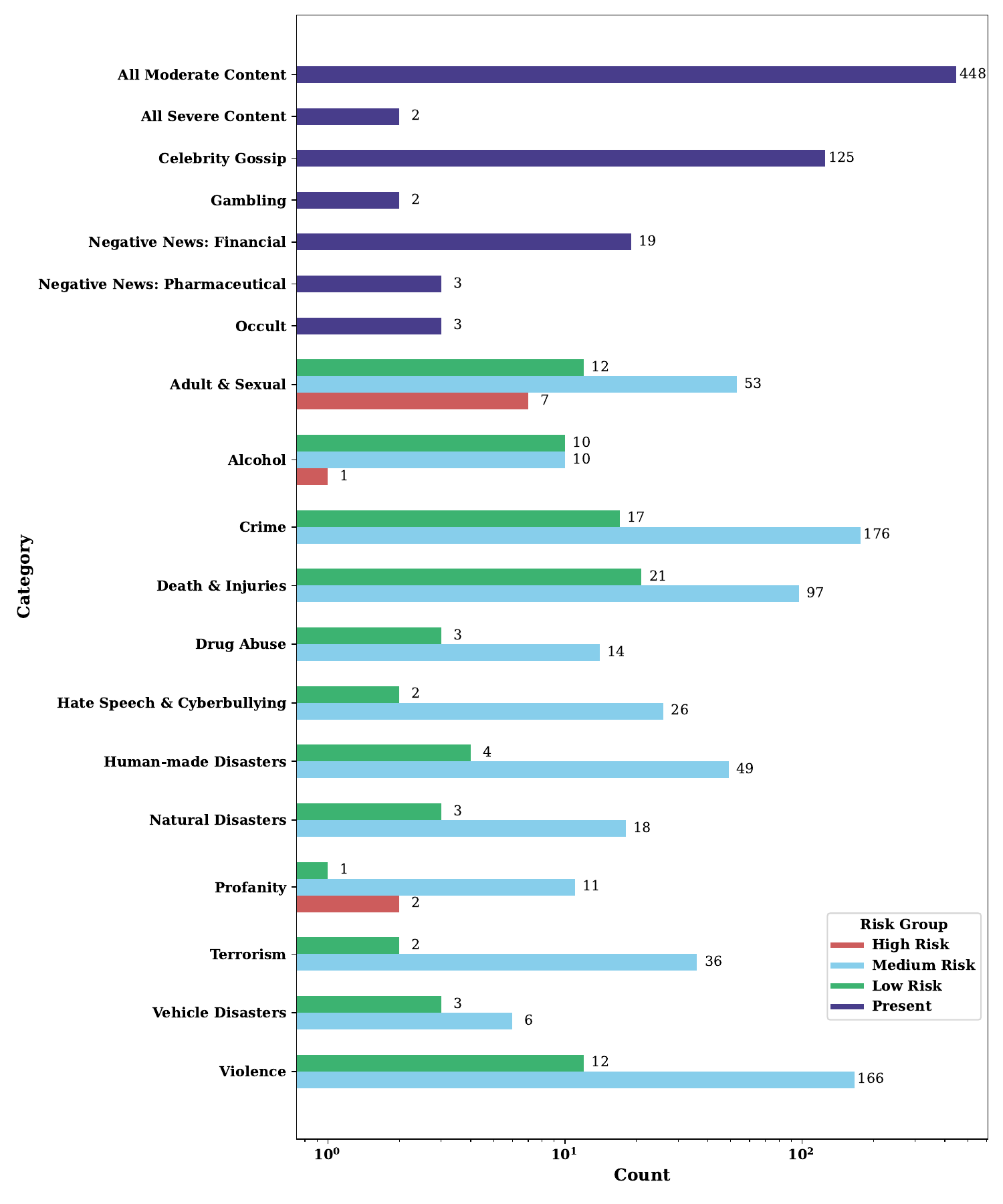}
\caption{DV Specific Classifications}
\label{fig:dv_categories_appendix}
\end{figure}

We collected data on 19 specific categories as described in Section ~\ref{sec:dv_categories_desc}. We chart the frequency of all 19 categories in Figure ~\ref{fig:dv_categories_appendix}. Similar to Oracle, the ``Crime'' and ``Death \& Injury'' are among the most common categories, with over 100 occurrences each. Similar to IAS, the ``Violence'' category is very common with 178 occurrences. The only other category with over 100 total occurrences is the ``Celebrity Gossip'' category (which does not exist under the other classification systems). DV rarely rates content as unsafe with ``High Risk'' as it only occurs 10 times in total (7 for ``Adult \& Sexual'', 2 for ``Profanity'' and, 1 for ``Alcohol''). Oddly, the ``Low Risk'' level is also not found frequently. It is only used 90 times, which is quite low when considering that the ``Medium Risk'' level is found 662 times across all categories.

\section{Appendix: Categorization of Domains}\label{sec:appendix_desc}

We classify each domain in our sample into a category of news so we can see if patterns exist in the classifications of similar domains. We take a simple approach to classifying each domain. We search for the website on Google and then record the blurb that is generated to describe the website. For each domain, we have copied that text (or a portion of that text) into the ``Description'' column of Table ~\ref{tab:website_categories}. For a few domains, there is no generated blurb. For these domains, we visit the about page on the website and copy the description from there. These domains are marked with an asterisk in Table ~\ref{tab:website_categories}. 

Once obtaining the descriptions, we can easily classify each domain. Any description that contains the words ``breaking... news'' is classified into the ``Breaking News'' category. If the description states it is the online format for a newspaper (e.g., azcentral.com) or if the descriptions state their focus as news for a specific global or local market (e.g., cnn.com or jsonline.com), then we classify them as ``General News''. Any domain that has a description with a long list of unrelated topics covered (e.g., theatlantic.com or usatoday.com) is also classified as ``General News''. In the main results section we combine domains which are classified as ``Breaking News'' or ``General News'' into one category because we see that the content on these domains is very similar. Domains with a description that clearly states their focus is in one area (e.g., wired.com and jalopnik.com) are classified into a category that is named to align with that focus (e.g., ``Technology'' and ``Cars''). For most of these domains which focus on a specific area, they are in a category of their own. The exceptions are the ten domains in the ``Finance/Business'' category and the five domains in the ``Entertainment'' category. We merge the business and finance focused domains because there are several which would fall under both categories anyway (e.g., bloomberg.com and cnbc.com). The domains classified as ``Entertainment'' either explicitly state that they focus on entertainment news (e.g. variety.com), or we consider their focus to be purely for entertainment (e.g. kotaku.com focuses on video gaming news). 



\onecolumn
\begin{longtable}{|p{0.17\textwidth}|p{0.63\textwidth}|p{0.14\textwidth}|}
\hline
\textbf{Domain} & \textbf{Description} & \textbf{Categorization} \\
\hline
\endfirsthead
\hline
\textbf{Domain} & \textbf{Description} & \textbf{Categorization} \\
\hline
\endhead
azcentral.com & The digital home of The Arizona Republic newspaper, with breaking news and in-depth coverage of sports, things to do, travel and opinions & Breaking News \\ \hline
barrons.com & Barron's is a leading source of financial news & Finance/Business \\ \hline
bizjournals.com & The Business Journals features local business news from 40-plus cities across the nation & Finance/Business \\ \hline
bloomberg.com & Bloomberg delivers business and markets news, data, analysis, and video to the world & Finance/Business \\ \hline
businessinsider.com & Business Insider tells the global tech, finance, stock market, media, economy, lifestyle, real estate, AI and innovative stories you want to know & Finance/Business \\ \hline
buzzfeed.com & BuzzFeed has breaking news, vital journalism, quizzes, videos, celeb news, Tasty food videos, recipes, DIY hacks, and all the trending buzz & Breaking News \\ \hline
cnbc.com & CNBC is the world leader in business news and real-time financial market coverage & Finance/Business \\ \hline
cnn.com & CNN is the world leader in news and information and seeks to inform, engage and empower the world. & General News \\ \hline
dailymail.co.uk & Get the latest breaking news, showbiz \& celebrity photos, sport news \& rumours, viral videos and top stories & Breaking News \\ \hline
dailytelegraph.com.au & News and Breaking News Headlines Online including Latest News from Australia and the World. & Breaking News \\ \hline
desmoinesregister.com & The Des Moines Register is the number one source for Des Moines and Iowa breaking, politics, business, agriculture, Iowa sports and entertainment news. & Breaking News \\ \hline
detroitnews.com & Your first source for breaking news, local in-depth reporting, and analysis of events important to Detroit and Michigan & Breaking News \\ \hline
dispatch.com & The Columbus Dispatch is the number one source for Columbus and Ohio breaking politics, business, obituaries, Ohio sports and entertainment news. & Breaking News \\ \hline
economist.com & Get in-depth global news and analysis. Our coverage spans world politics, business, tech, culture and more & General News \\ \hline
euronews.com & European and international latest breaking news, economic news, business news and more. & Breaking News \\ \hline
forbes.com & Forbes is a global media company, focusing on business, investing, technology, entrepreneurship, leadership, and lifestyle. & Finance/Business \\ \hline
ft.com & News, analysis and opinion from the Financial Times on the latest in markets, economics and politics. & Finance/Business \\ \hline
hollywoodreporter.com & Movie news, TV news, awards news, lifestyle news, business news and more & Entertainment \\ \hline
huffpost.com & Read the latest headlines, news stories, and opinion from Politics, Entertainment, Life, Perspectives, and more. & General News \\ \hline
investors.com & Perform stock investment research with our IBD research tools to help investment strategies. & Finance/Business \\ \hline
jalopnik.com* & Jalopnik is a news and opinion website about cars & Cars \\ \hline
jsonline.com & Milwaukee and Wisconsin news, sports, business, opinion, entertainment, lifestyle and investigative reporting & General News \\ \hline
kotaku.com & Gaming Reviews, News, Tips and More. & Entertainment \\ \hline
law.com & The premier global source for trusted and timely legal news, analysis and data. & Legal News \\ \hline
marketwatch.com & MarketWatch provides the latest stock market, financial and business news & Finance/Business \\ \hline
mashable.com & Mashable is a global, multi-platform media and entertainment company. & Entertainment \\ \hline
metro.co.uk & Metro.co.uk: News, Sport, Showbiz, Celebrities from Metro. & General News \\ \hline
msnbc.com & MSNBC breaking news and the latest news for today & Breaking News \\ \hline
news.com.au & The Latest on the news that matters to you - sport, entertainment, finance and politics. & General News \\ \hline
newyorker.com & The New Yorker is an American magazine featuring journalism, commentary, criticism, essays, fiction, satire, cartoons, and poetry. & General News \\ \hline
nymag.com & New York Magazine obsessively chronicles the ideas, people, and cultural events that are forever reshaping our world. & General News \\ \hline
nytimes.com & Live news, investigations, opinion, photos and video by the journalists of The New York Times from more than 150 countries around the world. & General News \\ \hline
qz.com & Quartz is a guide to the new global economy for people who are excited by change. We cover business, finance, economics & Finance/Business \\ \hline
realclearpolitics.com & RealClearPolitics (RCP) is an independent, non-partisan media company that is the trusted source for the best news, analysis and commentary. & Political News \\ \hline
reuters.com & Find latest news from every corner of the globe at Reuters.com, your online source for breaking international news coverage. & Breaking News \\ \hline
telegraph.co.uk & A British daily broadsheet newspaper & General News \\ \hline
theatlantic.com & The Atlantic covers news, politics, culture, technology, health, and more, through its articles, podcasts, videos, and flagship magazine. & General News \\ \hline
thedailybeast.com* & The Daily Beast delivers award-winning original reporting, fact-informed analysis, and sharp opinion in the arena of politics, pop-culture, and power and reaches more than 1 million readers a day. & General News \\ \hline
theglobeandmail.com & The Globe and Mail offers the most authoritative news in Canada, featuring national and international news. & General News \\ \hline
theguardian.com & Latest US news, world news, sports, business, opinion, analysis and reviews from the Guardian, the world's leading liberal voice. & General News \\ \hline
theroot.com* & Black News and Black Views with a Whole Lotta Attitude & General News \\ \hline
thesun.co.uk & Breaking headlines and latest news from the UK and the World & General News \\ \hline
theweek.com & Concise, twice-daily news digests curated by our editors. Distilled from dozens of the world's most trusted news sources & General News \\ \hline
time.com & Breaking news and analysis from TIME.com. Politics, world news, photos, video, tech reviews, health, science and entertainment news. & Breaking News \\ \hline
usatoday.com & Current national and local news, sports, entertainment, finance, technology, and more & General News \\ \hline
vanityfair.com & Vanity Fair is an American monthly magazine of popular culture, fashion, and current affairs & Fashion \\ \hline
variety.com & Entertainment news, film reviews, awards, film festivals, box office, entertainment industry conferences. & Entertainment \\ \hline
vox.com & Vox is a general interest news site for the 21st century. & General News \\ \hline
vulture.com & Vulture is a New York Magazine site providing continuous entertainment news covering TV, movies, music, art, books, theater, comedy, podcasts, celebrities, & Entertainment \\ \hline
washingtonpost.com & Breaking news, live coverage, investigations, analysis, video, photos and opinions & Breaking News \\ \hline
wired.com & Focuses on how emerging technologies affect culture, the economy, and politics & Technology \\ \hline
\caption{Categorization of Domains\\
\footnotesize Note: * means no Google blurb was available, so description is from the website itself}
\label{tab:website_categories}
\end{longtable}